\documentclass[11pt]{article}
\usepackage{chaithanya}
\usepackage{graphicx}
\usetikzlibrary{arrows.meta}

\newcommand{\cK}{\mathcal{K}}
\newcommand{\bz}{\boldsymbol{z}}

\title{Efficient classical simulation of Fermi-Hubbard models \\ with Kramers symmetry}
\author{
Chaithanya Rayudu\thanks{University of Cambridge, sscr2@cam.ac.uk}
\and
Jun Takahashi\thanks{University of Tokyo, juntakahashi@issp.u-tokyo.ac.jp}
\and
Ojas Parekh\thanks{Sandia National Laboratories, \{odparek, azhao\}@sandia.gov}
\and
Andrew Zhao\footnotemark[3]
}
\date{\today}

\begin{document}

\maketitle

\begin{abstract}
    Quantum Monte Carlo methods are widely used to solve quantum many-body problems. However, rigorous guarantees for their performance are poorly understood, especially for simulating interacting fermions. In this work, we develop a quantum Monte Carlo algorithm for a broad family of fermionic systems exhibiting Kramers time-reversal symmetry, and prove that their partition functions can be estimated in polynomial time. This family, which contains many important classes of Hubbard-type models, is known to both avoid the sign problem and obey a Lee-Yang zero-freeness theorem. Thus our results suggest that these two properties together may enable efficient classical algorithms for quantum many-body systems more broadly.
\end{abstract}

\tableofcontents

\section{Introduction}

Simulating quantum many-body systems is at the heart of modern computational physics and chemistry. 
A powerful tool for this task is the quantum Monte Carlo method, a suite of classical probabilistic algorithms which has enjoyed a long record of empirical success~\cite{foulkes2001quantum,assaad2008world,sandvik2010computational}. 
Despite their widespread adoption, however, rigorous guarantees for such methods are relatively sparse~\cite{crosson2025classicalsimulation,crosson2021rapid,takahashi2024rapidly,rayudu2025xy}. 
Moreover, such analyses are mostly confined to quantum spin models, whereas fermionic systems can provide a richer description of correlated quantum matter~\cite{arovas2022hubbard}. 
A theoretical understanding for the complexity of fermionic quantum Monte Carlo is therefore largely unexplored, even when the sign problem is absent~\cite{li2019sign}. 
Even beyond the scope of quantum Monte Carlo, only a limited number of strongly interacting fermionic systems are known to admit provably efficient classical algorithms for computing equilibrium properties~\cite{ramkumar2026high,jiang2026impurity,arunachalam2026quantum,schmidhuber2026polynomial}.

Understanding the structural properties that enable efficient simulation is central to expanding this frontier. 
For example, systems free of a sign problem can be treated straightforwardly with quantum Monte Carlo; however, this does not by itself guarantee the efficiency of the approach~\cite{hastings2013obstructions,marvian2019computational}. 
On the other hand, recent advances in \emph{quantum} algorithms~\cite{rayudu2026gap,bravyi2026heisenberg} have shown that the ground-energy problem can be efficiently solved, using a quantum computer, for systems obeying a Lee-Yang zero-freeness theorem~\cite{lee1952statistical,asano1970theorems,suzuki1971zeros}.  
For spin systems, this class includes (as a strict subset) Hamiltonians for which efficient classical algorithms have also been established, including stoquastic XY models~\cite{bravyi2017ferromagnets,rayudu2025xy} and, very recently, ferromagnetic Heisenberg models\footnote{By a standard local basis change, this includes bipartite antiferromagnetic Heisenberg models.}~\cite{chen2026fpras}. 
Notably, these examples are all stoquastic, suggesting a potential connection between the Lee-Yang property, the sign problem, and efficient classical simulation.

For fermionic systems, a version of the Lee-Yang theorem was recently discovered, covering a wide range of strongly interacting Fermi-Hubbard models~\cite{rayudu2026fermions}. 
Hubbard models provide a central paradigm for understanding electron correlation in materials, exhibiting rich phase diagrams that encompass Mott metal-insulator transitions, unconventional superconductivity, and more~\cite{arovas2022hubbard}. 
Analogous to the spin case, this class also includes many Hamiltonians which are sign-problem-free by certain fermionic quantum Monte Carlo representations~\cite{li2019sign}. 
However, unlike the situation for spins, the computational efficiency of this intersection is essentially unknown. 
This motivates us to study this intersection, both because of the importance of the Hubbard models within, and also to further explore this connection between the freeness of partition function zeros and of the sign problem.

We initiate this study here, giving a polynomial-time classical algorithm for computing the partition function of a broad family of Hamiltonians lying in this intersection. 
Specifically, our algorithmic guarantees apply to fermionic Lee-Yang Hamiltonians obeying Kramers time-reversal symmetry~\cite{sakurai2020modern}.

\begin{theorem*}[Informal]
For any fermionic Hamiltonian $H$ on $2N$ modes with \emph{attractive interactions} and \emph{Kramers symmetry} (\cref{def:kramers}), there is a classical randomized algorithm that estimates the partition function $\ZZ = \Tr e^{-\beta H}$ to relative error $\varepsilon$ in time $\mathrm{poly}(N,\beta,1/\varepsilon)$. Consequently, we can efficiently estimate the free energy and the ground energy of $H$ to any $1/\mathrm{poly}(N)$ additive error.
\end{theorem*}
Importantly, there are no restrictions on the strength or extensivity of the fermionic interactions. 
Canonical examples in this family include~\cite{assaad2008world}:
\begin{itemize}
    \item Attractive Fermi-Hubbard models with real hopping amplitudes on any graph;
    \item Repulsive Fermi-Hubbard models at half filling, with real hopping amplitudes on any bipartite graph (by a staggered particle-hole transformation).
\end{itemize}
For other notable examples, see \cite{wu2005sufficient}.

Kramers symmetry is well known to provide a positive-sign expansion of the partition function via the Hubbard-Stratonovich transformation~\cite{wu2005sufficient}, a cornerstone for determinant quantum Monte Carlo~\cite{scalapino1981method,blankenbecler1981monte,hirsch1983discrete}. 
However, our algorithm does not rely on Hubbard-Stratonovich. 
Instead, it more closely resembles the interaction expansion from continuous-time quantum Monte Carlo~\cite{rubtsov2005continuous}, although we work in discrete time. 
Our key insight is the development of a determinantal worm Markov chain, analogous to the worm algorithm for spins~\cite{prokofev1998exact,PROKOFEV1998253,sandvik1999stochastic} and systematically extending beyond ideas proposed previously for fermions~\cite{burovski2006fermi,goulko2010thermodynamics,wang2014fermionic}. 
We show that Kramers symmetry preserves sign positivity of the interaction expansion not only in the original configuration space, but also in this expanded worm-configuration space. 
Then, following recent breakthroughs on the fast mixing of Markov chains over certain log-concave and related measures~\cite{anari2024logconcave,chen2026fpras}, we prove that Kramers symmetry provides precisely the necessary ingredients to establish polynomial-time convergence of our quantum Monte Carlo algorithm.

\subsection{Main results}\label{sec:setup}

Here, we summarize our main results formally. Consider a fermionic system of an even number $2N$ of modes, described by the Fock space $\cF = \bigoplus_{k=0}^{2N}{\bigwedge}^k \C^{2N}$. Let $\ba=(a_1,\dots,a_{2N})^\T$ be the vector of fermionic annihilation operators. For a vector $v\in\C^{2N}$, let $a(v)\coloneqq v^\dagger\ba$. If $v$ is a unit vector, then $a(v)$ is the annihilation operator of an excitation in that mode $v$, and $n(v)\coloneqq a(v)^\dagger a(v)=\ba^\dagger vv^\dagger\ba$ is the corresponding number operator.

Kramers symmetry is defined with respect to the notion of \emph{antiunitary} operators. An operator $\Theta$ on a complex inner product space is antiunitary if
\begin{align}
    \Theta(\alpha u+\beta v)=\bar\alpha\,\Theta u+\bar\beta\,\Theta v \quad \text{and} \quad \langle\Theta u,\Theta v\rangle=\langle v,u\rangle
\end{align}
for all vectors $u,v$ and all $\alpha,\beta\in\C$, where we use $\bar{\alpha}$ to denote the complex conjugate of $\alpha$. Formally, we define the attractive Kramers class of interacting fermionic Hamiltonians as follows. 

\begin{definition}[Attractive Kramers class]\label{def:kramers}
A Hamiltonian $H$ on a Fock space $\cF$ on $2N$ modes is in the \emph{attractive Kramers class} if there exists an antiunitary operator $\Theta$ on $\C^{2N}$ with $\Theta^2=-\I$, a Hermitian matrix $h\in\C^{2N\times2N}$ with $\Theta h\Theta^{-1}=h$, unit vectors $v_1,\dots,v_r\in\C^{2N}$, and nonnegative numbers $U_1,\dots,U_r\ge0$ such that, with $w_i\coloneqq\Theta v_i$,
\begin{align}\label{eq:H}
H=H_0+\sum_{i=1}^rV_i,\quad\text{where}\quad H_0\coloneqq\ba^\dagger h\ba\quad\text{and}\quad V_i\coloneqq-U_i\,n(v_i)\,n(w_i).
\end{align}
\end{definition}

For an inverse temperature $\beta>0$, the partition function of $H$ is $\cZ\coloneqq\Tr e^{-\beta H}$ and the free energy is $F_\beta\coloneqq-\beta^{-1}\log\cZ$. We let $E_0$ denote the smallest eigenvalue of $H$ and define $\Lambda\coloneqq\|h\|_1+\sum_iU_i$. Our main result is the following theorem.

\begin{theorem}[Partition function]\label{thm:Z}
For any Hamiltonian $H$ in the attractive Kramers class, there is a randomized classical algorithm that, with probability at least $3/4$, outputs an estimate $\hat\cZ$ with $(1-\varepsilon)\cZ\le\hat\cZ\le(1+\varepsilon)\cZ$ in time $\mathrm{poly}(N,r,\beta,\Lambda,1/\varepsilon)$. 
\end{theorem}
\noindent Consequently, we can also efficiently estimate the free energy $F_\beta$ and the ground energy $E_0$.
\begin{corollary}[Free energy and ground energy]\label{cor:energies}
For any Hamiltonian $H$ in the attractive Kramers class, there are randomized classical algorithms that, with probability at least $3/4$, estimate $F_\beta$ to additive error $\varepsilon$ in time $\mathrm{poly}(N,r,\beta,\Lambda,1/\varepsilon)$, and $E_0$ to additive error $\varepsilon$ in time $\mathrm{poly}(N,r,\Lambda,1/\varepsilon)$.
\end{corollary}

\subsection{Related work}\label{sec:related_work}

\paragraph{Fermionic quantum Monte Carlo.}
The prominence of quantum Monte Carlo for simulating interacting fermions can hardly be understated, with many variations tailored for whether the system is \emph{ab initio}~\cite{foulkes2001quantum,motta2018ab}, is a lattice model~\cite{assaad2008world,li2019sign}, hosts impurities~\cite{gull2011continuous}, and so forth. 
The base formulation of our algorithm resembles continuous-time quantum Monte Carlo in the interaction expansion (CT-INT)~\cite{rubtsov2005continuous}, although we use a Trotterized discrete-time version. 
To our knowledge, the general principle behind the attractive Kramers class avoiding the sign problem in this particular representation has not been fully explored. 
Instead, it is typically through the Hubbard-Stratonovich transformation that the lack of sign problem for Kramers-invariant systems has been recognized~\cite{wu2005sufficient}. 
We clarify the relation between this condition and our \cref{def:kramers} in \cref{app:containment}. 
Other mechanisms beyond Kramers symmetry also provide positive-sign fermionic quantum Monte Carlo frameworks; see \cite{li2019sign} for a review.

Nonetheless, special cases of Kramers symmetry leading to positivity in CT-INT have been observed in the literature previously. 
In~\cite{bourovski2004truncated} it was recognized that the configuration weights take the form $(-U)^p \det(A_p^{\uparrow}) \det(A_p^{\downarrow})$ where $A_p^\sigma$ are certain Green's functions of the model. This is always nonnegative for attractive $U$ and $A_p^{\uparrow} = A_p^{\downarrow}$ real, which is indeed the case for attractive Kramers-invariant Hubbard models. 
They also point out that this holds for repulsive models at half filling on bipartite lattices. 
The comprehensive review by \cite{gull2011continuous} also discusses this phenomenon, although their treatment concerns impurity models which may additionally exhibit different, model-specific positivity mechanisms~\cite{yoo2005sign}.

\paragraph{Worm techniques for Markov chains.}
A prominent feature of Markov chain Monte Carlo techniques is that drastic update rules are allowed even if they do not necessarily directly correspond to physical processes. As long as the chain still converges to the target distribution and results in a faster convergence, a wild update method is even encouraged. 
The worm (also known as {\it directed loops}) method can be understood as a broad family of update rules where one needs to sample from a configuration representation with local constraints. 
Examples range from constrained classical models  \cite{prokofev2001worm} such as dimer models \cite{dimerworm}, vertex models \cite{vertexworm}, and loop models \cite{loopworm} to quantum models including spin \cite{spinwormProkofev,spinwormAnders}, bosonic \cite{PROKOFEV1998253,boninsegni2006worm}, and fermionic models \cite{burovski2006fermi,burovski2006critical,goulko2010thermodynamics,wang2014fermionic}. 

The diverse scope of the worm algorithm reflects the curious commonality in many highly-constrained configurations: they tend to allow {\it two} violations to be created. The idea of the worm method is then to let the two violations (also referred to as the worm head and tail) to engage in random walks, until they meet again and pair-annihilate, ending up in another constrained configuration. 
This could be seen as a condensed-matter physics rediscovery of the Jerrum-Sinclair chain for sampling perfect matchings using random walks of pair-created defects \cite{broder1986hard,jerrum1989approximating}. 
For physical worm algorithms, the worm heads can be understood as an auxiliary source term in the Hamiltonian, and enables direct and often more efficient sampling of correlation functions \cite{prokofev1998exact,boninsegni2006worm,sandvik2010computational}.

\paragraph{Efficient classical algorithms for quantum many-body problems.}
Here we give a nonexhaustive overview of classical algorithms for computing ground energies or partition functions in polynomial time. 
Although we do not review quantum algorithmic developments, mapping out the classical landscape is important for ruling out regimes where superpolynomial quantum advantage cannot be obtained.

For spin systems, the ground states of one-dimensional gapped chains can be rigorously treated via matrix product states~\cite{landau2015polynomial,arad2017rigorous}, validating their empirical success~\cite{white1992density}. 
Beyond one dimension, ferromagnetic transverse-field Ising models were shown to admit efficient partition function estimation~\cite{bravyi2015monte} by reducing to the classical case~\cite{jerrum1993ising}. 
This was later subsumed by similar guarantees for ferromagnetic XY models~\cite{bravyi2017ferromagnets} via reduction to sampling matchings~\cite{jerrum1989approximating}. Subsequent work proved that a genuine quantum Monte Carlo algorithm, implemented in practice for such XY systems, also features a fast mixing Markov chain with significantly improved bounds~\cite{rayudu2025xy}. 
This is in line with a body of work showing not only that a quantum spin system can be efficiently simulated, but that a particular instantiation of quantum Monte Carlo accomplishes this task~\cite{crosson2025classicalsimulation,crosson2021rapid,takahashi2024rapidly}. 
The models covered by these works include quantum Ising, XY, and Heisenberg models within a variety of parameter and temperature regimes.
In particular, the ferromagnetic Heisenberg model (and hence also the antiferromagnetic bipartite Heisenberg model) was recently shown to admit a fast mixing Markov chain for sampling from its partition function~\cite{chen2026fpras}. 
This result not only solves a major open problem in quantum many-body theory, but has applications to a remarkable number of other fields.

For fermionic systems, the above techniques can be applied whenever there exists a mapping from the fermions to a qualifying local spin system. However this is rather restrictive, especially beyond one dimension. 
The works of \cite{jiang2026impurity,arunachalam2026quantum} considered the quantum impurity model~\cite{bravyi2017impurity}, a strongly interacting system wherein the interactions are localized to a constant-size region. 
Their algorithms are efficient for ground energies and partition functions with polynomial inverse-temperature dependence. 
Besides this, however, known results are mostly confined to high-temperature thermal properties; for example, on bounded-degree interaction graphs above a critical temperature~\cite{ramkumar2026high}, in perturbative regimes~\cite{chen2025cumulant,han2026weakly}, or for random Hamiltonian models~\cite{schmidhuber2026polynomial}.

\subsection{Discussion}

We have presented, to our knowledge, the first polynomial-time guarantees for fermionic quantum Monte Carlo. 
Our algorithm applies to attractive fermionic systems with Kramers symmetry, a subclass of fermionic Lee-Yang Hamiltonians~\cite{rayudu2026fermions} that exhibit no sign problem~\cite{wu2005sufficient}. 
These results therefore rigorously validate decades of numerical success of methods such as determinant quantum Monte Carlo and its variants applied to such systems~\cite{hirsch1985two,white1989numerical,assaad2008world,li2019sign}. 

Notably, our algorithm is not itself a variant of determinant quantum Monte Carlo. 
Instead, we developed a determinantal worm Markov chain approach to a discrete-time version of continuous-time quantum Monte Carlo~\cite{rubtsov2005continuous}. 
While worm-type updates have been utilized in fermionic quantum Monte Carlo before~\cite{burovski2006fermi,goulko2010thermodynamics,wang2014fermionic}, our technique is systematically applicable to every Hamiltonian in the attractive Kramers class and has provable guarantees. 
Due to the power of such techniques, both theoretically and in practice, we hope that these ideas motivate further interest in worm Markov chains for fermionic systems.

Along with analogous results for spin systems~\cite{asano1970theorems,suzuki1971zeros,bravyi2017ferromagnets,rayudu2025xy,chen2026fpras}, our results also provide evidence that the intersection of Lee-Yang and stoquasticity may be a broader guiding principle for when quantum systems admit efficient classical simulation. For instance, even within the spin Lee-Yang Hamiltonians, the recent result of \cite{chen2026fpras} gives efficient classical simulation of Hamiltonians that are mappable to six-vertex models. But the spin Lee-Yang class also contains stoquastic (after a local basis rotation) Hamiltonians that are not mappable to these models. Currently, we do not have a provably efficient classical algorithm for such Hamiltonians. Similarly on the fermionic side, there are sign-problem-free fermionic Hamiltonians via Majorana reflection positivity \cite{wei2016majorana} that also have a Lee-Yang property \cite{rayudu2026fermions}. We expect all of these Hamiltonians to be efficiently classically simulable, and we leave that exploration for future research.

\section{Technical overview}\label{sec:overview}

\subsection{The quantum Monte Carlo algorithm}\label{sec:overview-qmc}
Our quantum Monte Carlo algorithm estimates the partition function in three steps. We first write the Trotter approximation of the partition function as a sum of nonnegative weights over classical configurations. Estimating this sum then reduces to sampling the configurations. Finally, we define a Markov chain that samples these configurations from a enlarged configuration space, where the enlarged space comes from adding a small pairing field to the Hamiltonian.

The first step uses standard Trotterization techniques. Split $e^{-\beta H}$ into $M$ Trotter steps of length $\delta\coloneqq\beta/M$ and write
\begin{align}
L \,\coloneqq\, CC^\dagger \quad \text{with} \quad C \,\coloneqq\, e^{-\delta H_0/2}\,\left(\I-\frac{\delta}{2}V_1\right)\cdots\left(\I-\frac{\delta}{2}V_r\right), \quad \text{so that} \quad L^M\approx e^{-\beta H}.
\end{align}
A polynomially large $M$ suffices to control the Trotter error between $L^M$ and $e^{-\beta H}$ (\cref{sec:proofs}). Each Trotter step $L$ contains every interaction factor $(\I-\frac{\delta}{2} V_i)$ twice, so $L^M$ contains $2rM$ such factors. We refer to these factors as \emph{insertions}, indexed by $\cK\coloneqq[M]\times[2r]$. Insertion $k$ has the form $\I+c_kP_k$, where $c_k\coloneqq\delta U_i/2$ and $P_k\coloneqq n(v_i)\,n(w_i)$ with interaction $i$ at that position. Expanding every insertion into its two terms, $\I$ and $c_kP_k$, turns $\Tr[L^M]$ into a sum over the sets $X\subseteq\cK$ of insertions at which $c_kP_k$ is chosen. Each term can be expressed as a free-fermion expectation value, hence by Wick's theorem~\cite{gaudin1960wick,bravyi2004lagrangian} we may write it as the determinant of two-point correlation functions of $H_0$. Specifically, in \cref{sec:green} we show that each term is proportional to a \emph{principal minor} of a global Green's function matrix $G\in\C^{2|\cK|\times2|\cK|}$. Each insertion contributes two rows and two columns to $G$, which we call its block row and its block column. The block row comes from the two annihilation operators in $P_k$, and the block column from the two creation operators. Writing $G[X]$ for the principal submatrix on the block rows and block columns in $X$, we get that
\begin{align}\label{eq:T_part_func_as_sum_of_dets}
\Tr\left[L^M\right] \,=\, \cZ_0\sum_{X\subseteq\cK}\det G[X],
\end{align}
where $\cZ_0$ is the efficiently computable partition function of $H_0$. This is the discrete-time analogue of the interaction expansion used in continuous-time quantum Monte Carlo \cite{rubtsov2005continuous}.

At first sight, there is no reason for the weights $\det G[X]$ to be nonnegative. Indeed, for generic hopping matrices and interactions, some of them are negative or complex, which is the infamous negative sign problem. Kramers symmetry, together with the attractive sign $U_i\ge0$ of the interactions, is what makes this expansion sign problem free. These two properties make $G$ \emph{quaternionic} (see \cref{sec:green}), which means that every $2\times2$ block of $G$ has the form $\bigl(\begin{smallmatrix}\alpha&-\bar\gamma\\ \gamma&\bar\alpha\end{smallmatrix}\bigr)$ of a quaternion. The eigenvalues of a square quaternionic matrix come in complex-conjugate pairs, and its real eigenvalues have even multiplicity, so its determinant is necessarily nonnegative. Moreover, square submatrices formed by whole block rows and block columns are again quaternionic, and hence every weight $\det G[X]$ is nonnegative. Hence the weights in the expansion \cref{eq:T_part_func_as_sum_of_dets} are nonnegative, and so we can define a probability distribution $\mu(X) \propto \det G[X]$ over the subsets $X \subseteq \cK$. Then, to estimate the sum $\Tr[L^M]$, we use the standard reduction from counting to sampling \cite{jerrum1986random,jerrum1993ising}, thereby reducing the problem to sampling from subsets $X \subseteq \cK$ with probability proportional to $\mu(X)$. 

\subsection{A determinantal worm Markov chain}

Our main technical result is showing that we can efficiently sample from this distribution in polynomial time. We sample in an enlarged configuration space that includes all minors $\det G[A,B]$ with $|A|=|B|$, not only the principal ones, where $G[A,B]$ keeps the block rows in $A$ and the block columns in $B$. 
These minors are also nonnegative because $G[A,B]$ is quaternionic, with the original principal minors corresponding to $A=B=X$. 
Relaxing the constraints of the configuration space in this way enables the Markov chain to explore the state space more freely, which is a well established philosophy~\cite{prokofev1998exact,prokofev2001worm}. 
We then bound the mixing time of the Markov chain defined over this extended configuration space.

\begin{figure}[t]
    \centering
    \begin{tikzpicture}[
      font=\small,
      selection/.style={black,line width=.6pt,-{Stealth[length=1.6mm]}},
      retained/.style={draw=black,fill=white,line width=1pt}
    ]
      \foreach \offset/\heading/\rows/\cols in
        {0/Valid configurations/X/X,6/Extended configurations/A/B} {
        \begin{scope}[xshift=\offset cm]
          \node[font=\normalsize] at (1.55,4.45) {\heading};
          \node at (1.55,3.85) {$\cols$: block columns};
          \node[rotate=90] at (-.85,1.55) {$\rows$: block rows};
    
          \draw[step=.62,densely dotted,line width=.4pt]
            (0,0) grid (3.1,3.1);
          \draw[line width=.6pt] (0,0) rectangle (3.1,3.1);
    
          \foreach \r in {1,3} {
            \draw[selection] (-.37,{.62*\r+.31}) -- (-.025,{.62*\r+.31});
          }
          \ifnum\offset=0
            \def\selectedcolumns{1,3}
          \else
            \def\selectedcolumns{0,3}
          \fi
    
          \foreach \c in \selectedcolumns {
            \draw[selection] ({.62*\c+.31},3.47) -- ({.62*\c+.31},3.125);
          }
    
          \foreach \r in {1,3} {
            \foreach \c in \selectedcolumns {
              \draw[retained] ({.62*\c},{.62*\r})
                rectangle ({.62*(\c+1)},{.62*(\r+1)});
              \draw[line width=.3pt]
                ({.62*\c+.31},{.62*\r}) -- ({.62*\c+.31},{.62*(\r+1)})
                ({.62*\c},{.62*\r+.31}) -- ({.62*(\c+1)},{.62*\r+.31});
            }
          }
        \end{scope}
      }
    
      \node[align=center] at (1.55,-.65)
        {$G[X]=G[X,X]$};
      \node[align=center] at (7.55,-.65)
        {$G[A,B],\quad |A|=|B|$};
    
      \draw[retained] (2.55,-1.65) rectangle (2.79,-1.41);
      \draw[line width=.3pt]
        (2.67,-1.65) -- (2.67,-1.41)
        (2.55,-1.53) -- (2.79,-1.53);
      \node[anchor=west] at (2.8,-1.53) {: Selected $2\times2$ blocks};
    \end{tikzpicture}
    \caption{Block rows and columns in the original and enlarged configuration spaces. Each side depicts the Green's function matrix $G$, partitioned into $2\times2$ blocks associated with interaction insertions. Arrows indicate selected block rows and columns. Each selected block row represents a pair annihilation $\Delta_i = a(w_i) a(v_i)$, and each selected block column represents a pair creation $\Delta_i^\dagger$. \textbf{Left:} In a valid configuration, block rows and columns are the same $X \subseteq \cK$, producing the principal submatrix $G[X]=G[X,X]$ with weight proportional to $\det G[X]$. Pair creation and annihilation therefore occur at the same insertions. \textbf{Right:} In the enlarged space, the block rows $A$ and block columns $B$ may differ, subject to $|A|=|B|$ so that $G[A,B]$ remains square. The rows in $A\setminus B$ contribute only pair annihilation, while the columns in $B\setminus A$ contribute only pair creation. These configurations have weight proportional to $\lambda^{|A\setminus B|}\det G[A,B]$, and conditioning on $A=B$ recovers the original distribution. Note that submatrices formed from $2 \times 2$ block rows and columns inherit the quaternionic structure of $G$, guaranteeing nonnegativity of all minors.}
    \label{fig:configurations}
\end{figure}
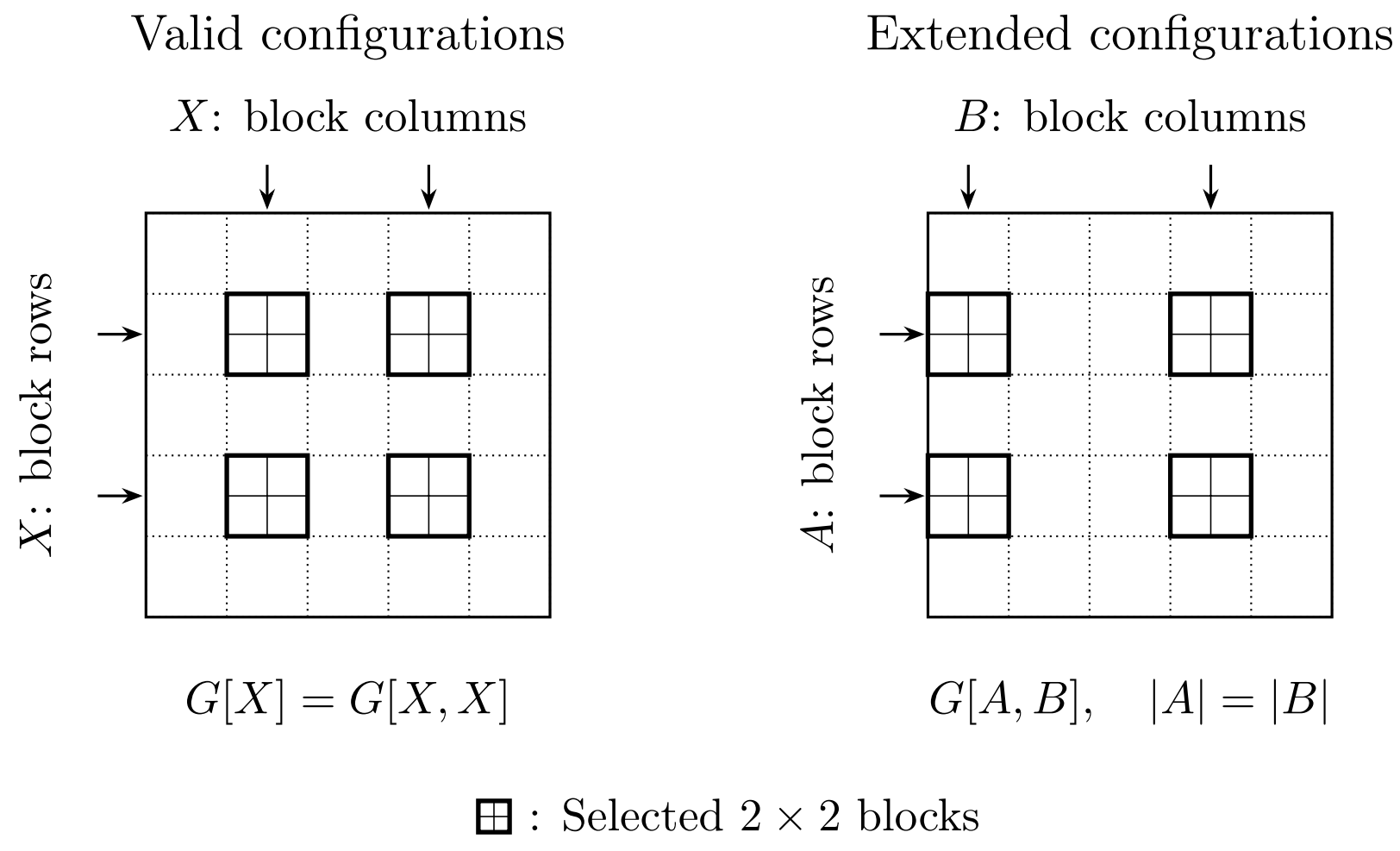

This enlargement has a simple physical interpretation: it amounts to adding a small pairing field to the Hamiltonian. For each interaction $i$, let $\Delta_i\coloneqq a(w_i)\,a(v_i)$, where $a(v)$ annihilates a fermion in the mode $v$. This operator annihilates a pair of fermions in the Kramers pair $v_i$ and $w_i=\Theta v_i$. At an insertion $k$ of interaction $i$, we write $\Delta_k$ for $\Delta_i$, so that $P_k=\Delta_k^\dagger\Delta_k$. We add the pairing field $\Delta_k+\Delta_k^\dagger$ as follows: every insertion $\I+c_kP_k$ of $L$ is replaced with $\I+c_kP_k+\sqrt{\lambda c_k}\,(\Delta_k+\Delta_k^\dagger)$, for some parameter $0<\lambda<1$. Write $L_\lambda$ for the resulting modified Trotter step. Each insertion in the expansion of $\Tr[L_\lambda^M]$ now has four options: $\I$, $P_k$, $\Delta_k$ and $\Delta_k^\dagger$. Let $A$ collect the insertions that annihilate a pair, that is, those contributing $P_k$ or $\Delta_k$. Similarly, let $B$ collect those that create a pair, contributing $P_k$ or $\Delta_k^\dagger$. Because $H_0$ and the $P_k$ conserve the fermion number, every term with $|A|\ne|B|$ vanishes. The remaining terms in the expansion of $\Tr[L_\lambda^M]$ are minors of $G$. More explicitly, since $\Delta_k$ contains only the annihilation operators of insertion $k$, and $\Delta_k^\dagger$ only its creation operators, Wick's theorem gives
\begin{align}\label{eq:pairing-overview}
\Tr\left[L_\lambda^M\right] \,=\, \cZ_0\sum_{\substack{A,B\subseteq\cK\\|A|=|B|}}\lambda^{|A\setminus B|}\det G[A,B].
\end{align}
Each of the $|A\setminus B|$ pair annihilations and $|B\setminus A|$ pair creations contributes a factor $\sqrt\lambda$, while the factor $\sqrt{c_k}$ in the pairing term matches the normalization of the block rows and block columns of $G$. Note that setting $A=B=X$, so that every insertion in $X$ contributes $P_k$, recovers exactly the terms of \cref{eq:T_part_func_as_sum_of_dets}.

Now, we define a Markov chain to sample from this enlarged space, where $(A,B)$ is sampled with probability proportional to $\lambda^{|A\setminus B|}\det G[A,B]$. Denote this distribution by $\pi_\lambda$. Each move first removes a block row or adds a block column uniformly at random, which is accepted with probability $1$ at the insertions in $A\setminus B$ and with probability $\lambda$ otherwise. We then restore $|A|=|B|$ by adding a block row or removing a block column, chosen with probability proportional to the determinant of the
resulting submatrix of G. We will efficiently obtain a sample with $A = B$ if those states are
visited sufficiently frequently by this Markov chain in the extended configuration space. 

We can quantify this by examining their total probability mass,
\begin{align}
\pi_\lambda(A=B) \,=\, \frac{\Tr\left[L^M\right]}{\Tr\left[L_\lambda^M\right]},
\end{align}
which is the ratio of the Trotterized partition functions without and with the pairing field. Intuitively, a weak pairing field only changes the partition function slightly; since each $\Delta_i+\Delta_i^\dagger$ has unit norm, a field of total strength of order $1/\beta$ changes $\Tr e^{-\beta H}$ by at most a constant factor. \Cref{app:valid} proves the necessary Trotterized version of this statement, which bounds $\pi_\lambda(A=B)$ from below by a constant whenever $\lambda$ is inverse-polynomially small. The insertions in $A\setminus B$ and $B\setminus A$ are thus defects of small weight, analogous to the odd-degree vertices in the Jerrum-Sinclair algorithm for the Ising model \cite{jerrum1993ising} and the ends of the worm in worm algorithms \cite{prokofev1998exact,prokofev2001worm,rayudu2025xy}.

So far, we used Kramers symmetry only to make the weights $\det G[A,B]$ nonnegative. In the next subsection, we explain how it also enables us to prove that the Markov chain mixes in polynomial time.

\subsection{Establishing fast mixing}

Now we provide an overview of the proof that the Markov chain defined above mixes in polynomial time. This implies that the quantum Monte Carlo algorithm described in \cref{sec:overview-qmc} is efficient for every Hamiltonian of \cref{def:kramers}. Our main technical tools are the rapid mixing of the basis-exchange chain for homogeneous strongly log-concave measures \cite{anari2024logconcave} and its extension \cite{chen2026fpras} to coefficient-wise tilts of such measures.

Recall that the Markov chain samples a configuration $(A,B)$ with probability proportional to $\lambda^{|A\setminus B|}\det G[A,B]$. We can represent each configuration $(A,B)$ by the monomial $\bz^A\by^{\cK\setminus B}$, with a variable $z_k$ for each block row in $A$ and a variable $y_k$ for each block column that is not in $B$. Since $|A|=|B|$, every such monomial has degree $|\cK|$. Our main technical observation is that, with $\lambda = 1$, Kramers symmetry makes the generating polynomial of the minors $\det G[A,B]$ strongly log-concave. In \cref{lem:Gpoly}, we show that this generating polynomial is the square root of a determinant,
\begin{align}
\sum_{\substack{A,B\subseteq\cK\\|A|=|B|}}\det G[A,B]\;\bz^{A}\,\by^{\cK\setminus B} \,=\, \sqrt{\det\left(D_{\by}+G^\dagger D_{\bz}G\right)}
\end{align}
for positive variables $\by$ and $\bz$, where $D_{\by}$ and $D_{\bz}$ are diagonal with blocks $y_k\I_2$ and $z_k\I_2$. This is a specific consequence of Kramers symmetry. Since the matrix $D_{\by}+G^\dagger D_{\bz}G$ is positive definite and depends linearly on $(\by,\bz)$, the logarithm of its determinant is concave, and for a multiaffine polynomial this implies strong log-concavity (\cref{lem:multiaffine}). The Markov chain to sample from homogeneous log-concave measures \cite{anari2024logconcave} is a basis-exchange chain. At each step, it removes a uniformly random element of the current set and then adds an element back, with probability proportional to the weight of the resulting set. Our chain makes the same kind of move on the monomials. Its first step removes one of the $|\cK|$ variables of the monomial, chosen uniformly at random. Deleting the block row $k\in A$ removes $z_k$, and adding the block column $k\notin B$ removes $y_k$. Its second step adds a variable back, with probability proportional to the coefficient of the resulting monomial. In the extreme case of strong pairing field, i.e., for $\lambda=1$, every removal is accepted, so our chain is the basis-exchange chain of the minors $\det G[A,B]$, and it mixes rapidly because the generating polynomial of the minors is homogeneous and strongly log-concave.

Adding the pairing field multiplies the weight of $(A,B)$ by $\lambda^{|A\setminus B|}$. This factor is a product over all the insertions in $\cK$. It contributes $\lambda$ for each insertion whose block row is in $A$ and whose block column is not in $B$, that is, for each $k$ at which both $z_k$ and $y_k$ appear in the monomial. Multiplying the weights by such a factor need not preserve the log-concavity of the generating polynomial. Chen and Liu \cite{chen2026fpras} showed that the basis-exchange chain remains rapidly mixing when the weights are multiplied by such a factor, provided that the factor satisfies a local spectral curvature condition. For a factor that is a product over the insertions, this condition reduces to one condition for each insertion. In our case, this condition holds true for any $0<\lambda<1$. In \cref{sec:mixing}, we use this to show that the spectral gap of our Markov chain is at least $\lambda/(2|\cK|)$. Since $\lambda$ only needs to be inverse polynomially small and $|\cK|$ is only polynomially large, the chain mixes in polynomial time.

\section{Kramers symmetry and quaternionic matrices}\label{sec:kramers}

Let $\Theta$ be an antiunitary operator acting on the one-particle vector space $\C^{2N}$ of even dimension. Define $\Theta_\cF$ on the fock space $\cF$ by lifting $\Theta$ from the one-particle vector space as
\begin{gather}
    \Theta_\cF\,a(v)\,\Theta_\cF^{-1}=a(w) \quad \where \quad w = \Theta v,\\
    \text{and}\quad \Theta_\cF\,a^{\dagger}(v_1)\dots a^{\dagger}(v_r)\ket{\vac} = a^{\dagger}(w_1)\dots a^{\dagger}(w_r)\ket{\vac} \quad \where \quad w_i = \Theta v_i \quad \text{for all } i.
\end{gather}
If $\Theta$ commutes with a hopping matrix $h \in \C^{2N \times 2N}$, then the lift $\Theta_\cF$ commutes with $H_0 = \ba^\dagger h\ba$. Also, for every $w_i = \Theta v_i$ where $v_i,w_i \in \C^{2N}$, $\Theta_\cF$ exchanges $n(v_i)$ and $n(w_i)$. This implies $\Theta_\cF$ commutes with $n(v_i)n(w_i)$ when $w_i = \Theta v_i$. Combining the two commutation properties, we get that $\Theta_\cF$ commutes with any Hamiltonian $H$ of the form in \cref{def:kramers}. In other words, $\Theta_\cF$ is an antiunitary symmetry of $H$, also called as a time reversal symmetry in physics. Furthermore, we say $H$ has Kramers symmetry if $\Theta^2 = -\I$, as is the case in \cref{def:kramers}.

Kramers symmetry is useful in that the eigenvalues of every matrix commuting with $\Theta$ come in pairs, so the determinant of such a matrix is nonnegative. This is the reason for determinant quantum Monte Carlo to not have the sign problem for Hamiltonians with Kramers symmetry \cite{wu2005sufficient}. Our proofs also use this property. But the state space and the Markov chain in the our work is different from ones in the determinant quantum Monte Carlo (and its variants). To state some of the properties we use in our work about the Hamiltonians with Kramers symmetry, we choose a basis adapted to a given $\Theta$, in which the matrices commuting with $\Theta$ become quaternionic matrices.

\paragraph{Kramers basis.}
Taking $w=\Theta v$ in $\langle\Theta w,\Theta v\rangle=\langle v,w\rangle$ and using $\Theta^2=-\I$ by Kramers symmetry gives $\langle v,\Theta v\rangle=0$ for every $v$. Hence, for any unit vector $e_1$, the vectors $e_1$ and $\Theta e_1$ are orthonormal, and by the same identity the orthogonal complement of their span is invariant under $\Theta$. By induction, we can build an orthonormal basis $e_1,\Theta e_1,\dots,e_N,\Theta e_N$ of $\C^{2N}$, which we call a \emph{Kramers basis}. In the rest of the paper, we write all the vectors and matrices in this basis. Since $\Theta(\Theta e_b)=-e_b$, by antilinearity of $\Theta$, we get that
\begin{align}
\Theta v=-\bbJ_N\bar v\quad(v\in\C^{2N}),
\end{align}
in a Kramers basis where $\bar v$ is the entrywise complex conjugate of $v$ where, for every $p \in \Z_{+}$,
\begin{align}
\bbJ_p\coloneqq\I_p\otimes J,\quad J\coloneqq\begin{pmatrix}0&1\\-1&0\end{pmatrix}.
\end{align}
Note that $J^{-1} = J^{\T} = -J$ and therefore $\bbJ_p^{-1} = \bbJ^{\T}_p = -\bbJ_p$ for every $p \in \Z_{+}$.

\paragraph{Quaternionic matrices.}
A matrix $Q\in\C^{2N\times2N}$ commutes with $\Theta$ if and only if $Q\bbJ_N=\bbJ_N\bar Q$ in a Kramers basis. We call such matrices \emph{quaternionic}. More generally, we call a matrix $Q\in\C^{2p\times2q}$ \emph{quaternionic} if and only if
\begin{align}
    \label{def:quaternionic_matrices}
    Q\bbJ_q=\bbJ_p\bar Q.
\end{align}
Since $Q$ has even number of rows and columns, we can group rows into $p$ blocks where each block contains two consecutive rows, and similarly, the columns into $q$ blocks. Then the condition in \cref{def:quaternionic_matrices} is equivalent to having each of the $2\times 2$ blocks of $Q$ of the form
\begin{align}
\begin{pmatrix}\alpha&-\bar\gamma\\ \gamma&\bar\alpha\end{pmatrix},\quad\alpha,\gamma\in\mathbb C. 
\end{align}
Quaternionic matrices are closed under real linear combinations, products, inverses, conjugate transposes and under eliminating blocks of rows and columns. For $A\subseteq[p]$ and $B\subseteq[q]$, we write $Q[A,B]$ for the submatrix formed by the block rows in $A$ and the block columns in $B$, with $\det Q[\varnothing,\varnothing]\coloneqq1$, and, if $Q$ is square, $Q[A]\coloneqq Q[A,A]$ for the principal submatrix formed by the block rows and block columns in $A$.

\paragraph{Determinants and Pfaffians.}

The map $v\mapsto\bbJ_p\bar v$ is antiunitary, squares to $-\I$, and commutes with a quaternionic matrix $Q$. Hence eigenvalues of $Q$ come in complex-conjugate pairs, and every real eigenvalue has even multiplicity. This gives $\det Q\ge0$.

\begin{lemma}[Kramers determinants {\cite{wu2005sufficient}}]\label{lem:kdet}
If $Q\in\C^{2p\times2p}$ is quaternionic, then $\det Q\ge0$.
\end{lemma}

Furthermore, if $Q\in\C^{2p\times2p}$ is \emph{Hermitian}, using $\bbJ_p^{-1}=\bbJ_p^\T=-\bbJ_p$, we can show that $\bbJ_p Q$ is antisymmetric i.e.,
\begin{align}
    (\bbJ_p Q)^\T=Q^\T\bbJ_p^\T=\bbJ_p^{-1}Q\bbJ_p\bbJ_p^\T=\bbJ_p^{-1}Q=-\bbJ_p Q.
\end{align}
Therefore, the Pfaffian of $\bbJ_p Q$ is well defined and
\begin{align}
    \Pf(\bbJ_p Q)^2=\det(\bbJ_p Q)=\det Q.
\end{align}
The Pfaffian $\Pf(\bbJ_p Q)$ is just the Moore quaternion determinant of $Q$, viewed as a $p\times p$ quaternion matrix \cite{aslaksen2001quaternionic}. Moreover, if $Q$ is positive semidefinite,
\begin{align}
    \Pf(\bbJ_p Q)=\sqrt{\det Q}.\label{eq:P1}
\end{align}
\Cref{eq:P1} holds because, for positive semidefinite $Q$, the quaternion determinant is the product of the eigenvalues of the quaternion matrix, which are nonnegative \cite{alesker2003noncommutative}. 

\section{Fermionic quantum Monte Carlo}\label{sec:chain}

In this section, we explain the algorithm to estimate the partition function $\cZ = \Tr[e^{-\beta H}]$ via Trotterization, reduction from counting to sampling, and efficient sampling using a Markov chain. In \Cref{sec:trotter}, we define the Trotter step $L$ and expand $\Tr[L^M]$ into a sum over sets of insertions. \Cref{sec:green} then constructs the Green's function matrix $G$, shows that it is quaternionic, and identifies each term of the expansion of $\Tr[L^M]$ with $\cZ_0\det G[X]$. We end that section by recalling the standard counting to sampling reduction, whose details we give in \cref{app:counting}. In \cref{sec:chaindef}, we define a Markov chain to sample from an enlarged configuration space, where the enlarged space can be seen as adding on a small pairing field. We prove in \cref{app:valid} that the probability mass of the \emph{valid} configurations in this enlarged space is not too small. The proof that the Markov chain mixes rapidly is deferred to \cref{sec:mixing}. Finally, \cref{sec:proofs} combines everything together to show that we can efficiently estimate $\Tr[L^M]$ to the required precision.

\subsection{Trotterization and Wick's theorem}\label{sec:trotter}
Let $M$ be a positive integer, $\delta\coloneqq\beta/M$, $c_i\coloneqq\delta U_i/2$ and $P_i\coloneqq n(v_i)\,n(w_i)$, with $U_i$, $v_i$ and $w_i$ as in \cref{def:kramers}. With the free fermion Hamiltonian $H_0$ and the interaction terms $V_i = -U_i\,n(v_i)\,n(w_i)$ from \cref{def:kramers}, define the Trotter step
\begin{align}\label{eq:trotter}
L\coloneqq CC^\dagger, \quad \text{where} \quad C\coloneqq e^{-\delta H_0/2}\,\left(\I-\frac{\delta}{2}V_1\right)\cdots\left(\I-\frac{\delta}{2}V_r\right).
\end{align}
Since $H_0$ and $V_i$ are Hermitian, each Trotter step is
\begin{align}
L \,=\, e^{-\delta H_0/2}\,\left(\I-\frac{\delta}{2}V_{i_1}\right)\cdots\left(\I-\frac{\delta}{2}V_{i_{2r}}\right)\,e^{-\delta H_0/2},
\end{align}
where $(i_1,\dots,i_{2r})\coloneqq(1,\dots,r,r,\dots,1)$. Our goal is to estimate the Trotterized partition $\Tr[L^M]$, which in turn approximates partition function $\cZ$ to any relative precision for a polynomially large $M$ \cite{childs2021theory}. The product $L^M$ contains $2rM$ interaction factors, which we call \emph{insertions} and index them by $\cK\coloneqq[M]\times[2r]$. For $k=(m,q)\in\cK$ and $i=i_q$, insertion $k$ is the factor $\I-\frac{\delta}{2}V_i=\I+c_kP_k$, where
\begin{align}
c_k\,\coloneqq\,c_i \quad \text{and} \quad P_k\,\coloneqq\,n(v_i)\,n(w_i).
\end{align}
Expanding every insertion into its two terms, $\I$ and $c_kP_k$, and labeling the insertions at which $c_kP_k$ is chosen, we can write
\begin{align}\label{eq:expansion}
\Tr\left[L^M\right]=\sum_{X\subseteq\cK}f(X) \quad \text{with} \quad f(X)\coloneqq c^X\,\Tr\left[\,\prod_{m=1}^{M}e^{-\delta H_0/2}\left(\,\prod_{q=1}^{2r}P_{(m,q)}^{\mathbbm{1}[(m,q)\in X]}\,\right)e^{-\delta H_0/2}\,\right]
\end{align}
and $c^{X} \coloneqq \prod_{k\in X} c_k$. Later in this section, we show that the terms $f(X)$ are nonnegative, and we reduce the estimation of $\Tr[L^M]$ to sampling from a Markov chain on pairs $(A,B)$ of subsets of $\cK$, whose stationary distribution, conditioned on $A=B$, is proportional to $f(A)$.

Note that $f(X)$ in \cref{eq:expansion} is proportional to the expectation value, with respect to the Gibbs state of the free-fermion Hamiltonian $H_0$, of a product of imaginary-time evolved operators. Concretely, let
\begin{align}\label{eq:gibbs}
\langle O\rangle\coloneqq\frac{\displaystyle \Tr\left[e^{-\beta H_0}O\right]}{\displaystyle \cZ_0} \quad \text{with} \quad \cZ_0 \coloneqq \Tr\left[e^{-\beta H_0}\right] = \det\left(\I + e^{-\beta h}\right)
\end{align}
be the expectation value of an operator $O$.
For $k=(m,q)\in\cK$, let
\begin{align}
    \tau_k\coloneqq\left(M-m+\frac12\right)\delta \quad \text{and} \quad O(\tau_k)\coloneqq e^{\tau_kH_0}\,O\,e^{-\tau_kH_0}
\end{align}
for every operator $O$. Then, by arranging the elements of $X \subseteq \cK$ lexicographically, we get that
\begin{align}\label{eq:f-expectation}
f(X)=c^X\,\cZ_0\;\left\langle P_{k_1}(\tau_{k_1})\cdots P_{k_n}(\tau_{k_n})\right\rangle.
\end{align}
Note that $\tau_k$ for any $k = (m,q)$ depends only on the Totter slice number $m$. 
We can further divide each of the operators $P_k(\tau_k)$ for $k = (m,q)$ and $i=i_q$ into a product of imaginary-time evolved creation and annihilation operators as
\begin{gather}
P_k(\tau_k) \,=\, e^{\tau_kH_0}P_k\,e^{-\tau_kH_0}
\,=\, \left[a\left(e^{\tau_kh}v_i\right)\right]^\dagger\,a\left(e^{-\tau_kh}v_i\right)\,\left[a\left(e^{\tau_kh}w_i\right)\right]^\dagger\,a\left(e^{-\tau_kh}w_i\right), \\[10pt]
\text{since} \quad e^{tH_0}\left[a(x)\right]^\dagger e^{-tH_0}=\left[a\left(e^{th}x\right)\right]^\dagger, \quad e^{tH_0}a(x)\,e^{-tH_0}=a\left(e^{-th}x\right),\label{eq:heisenberg}
\end{gather}
for any $t \in \R$. We can now use Wick's theorem for the Gibbs state of $H_0$ to evaluate $f(X)$ efficiently for any given $X$ since each $f(X)$ is an expectation value of product of time-evolved creation and annihilation operators.

Let the one-particle density matrix of $H_0$ be
\begin{gather}
\rho\coloneqq\left(\I+e^{\beta h}\right)^{-1},\quad\text{so that}\quad\left(\I+e^{-\beta h}\right)^{-1}=\I-\rho, \label{eq:rho} \\[10pt]
\left\langle a(x)^\dagger a(y)\right\rangle =y^\dagger\rho\,x, \quad \text{and} \quad  \left\langle a(y)\,a(x)^\dagger\right\rangle =y^\dagger(\I-\rho)\,x.\label{eq:twopoint}
\end{gather}
Wick's theorem~\cite{gaudin1960wick,bravyi2004lagrangian} expresses the expectation value of any product of such pairs of creation and annihilation operators as a determinant of two-point functions.

\begin{lemma}[Wick's theorem]\label{lem:wick}
Let $x_1,\dots,x_n,y_1,\dots,y_n\in\C^{2N}$, and let $G$ be the $n\times n$ matrix with entries
\begin{align}
G_{ji}\coloneqq\begin{cases}\phantom{-}y_j^\dagger\rho\,x_{i}&\text{if }j\ge i,\\-y_j^\dagger(\I-\rho)\,x_{i}&\text{if }j<i.\end{cases}
\end{align}
Then
\begin{align}\label{eq:wick}
\left\langle a(x_1)^\dagger a(y_1)\,a(x_2)^\dagger a(y_2)\cdots a(x_n)^\dagger a(y_n)\right\rangle \,=\, \det G.
\end{align}
\end{lemma}
\noindent The columns and rows of the matrix $G$ are indexed by $\{x_1,\cdots,x_n\}$ and $\{y_1,\cdots,y_n\}$ respectively in that order. \Cref{lem:wick} is a special case of the Pfaffian form of Wick's theorem \cite{gaudin1960wick,bravyi2004lagrangian}, which states that $\langle b_1\cdots b_{2n}\rangle=\Pf(\Gamma)$ for any operators $b_1,\dots,b_{2n}$ that are linear combinations of creation and annihilation operators, where $\Gamma$ is the antisymmetric matrix with $\Gamma_{pq}\coloneqq\langle b_pb_q\rangle$ for $p<q$. We use this general form in \cref{sec:chaindef}.

\subsection{Green's function matrix} \label{sec:green}
Given that $f(X)$ is an expectation value of product of imaginary-time evolved operators, by \cref{eq:f-expectation}, and by \cref{lem:wick}, each term $f(X)$ in the expansion in \cref{eq:expansion} is proportional to the determinant of a matrix of two-point functions. We now show that all of these matrices are principal submatrices of a single matrix $G$ of size $2|\cK| \times 2|\cK|$.

For $k=(m,q)\in\cK$ and $i=i_q$, define
\begin{align}
R_k^{\pm}\coloneqq c_k^{1/4}\,e^{\pm\tau_kh}\,\left[\,v_i\ \ w_i\,\right]\in\C^{2N\times2},
\end{align}
so that the columns of $R_k^+$ and $R_k^-$ are, up to the factor $c_k^{1/4}$, the vectors of the imaginary-time evolved creation and annihilation operators, respectively, in the decomposition of $P_k(\tau_k)$. The factor $c_k^{1/4}$ includes the coupling constants in $G$: it multiplies each of the two pairs of creation and annihilation operators in the decomposition of $P_k(\tau_k)$ by $\sqrt{c_k}$, and hence their product by $c_k$. Let $G$ be the matrix that \cref{lem:wick} builds with the following $4|\cK|$ vectors,
\begin{align}
    \left[x_1\,\, \cdots \,\,x_{2|\cK|} \right] &=  \left[R_{(1,1)}^+\ \ R_{(1,2)}^+\cdots R_{(M,2r)}^+\right] \nonumber\\
    \quad \text{and} \quad  \left[y_1\,\, \cdots \,\,y_{2|\cK|} \right] &=  \left[R_{(1,1)}^-\ \ R_{(1,2)}^-\cdots R_{(M,2r)}^-\right].
\end{align}
Thus each insertion $k$ contributes two rows and two columns to $G$, which we call its block row and its block column. The block row comes from the two annihilation operators in $P_k$, and the block column from the two creation operators. When $G$ is grouped into $2\times2$ blocks indexed by $\cK\times\cK$, the block in position $(k,l)$ is
\begin{align}\label{eq:G}
G_{kl}=
\begin{cases}
\phantom{-}\left(R_k^-\right)^\dagger\,\rho\,R_l^+&\text{if }k>l,\\[6pt]
\sqrt{c_k}\begin{pmatrix}v_i^\dagger\rho\,v_i&-v_i^\dagger(\I-\rho)\,w_i\\[6pt] w_i^\dagger\rho\,v_i&w_i^\dagger\rho\,w_i\end{pmatrix}&\text{if }k=l=(m,q),\ i=i_q,\\[20pt]
-\left(R_k^-\right)^\dagger\,(\I-\rho)\,R_l^+&\text{if }k<l.
\end{cases}
\end{align}
\begin{lemma}\label{lem:Gquat}
The matrix $G$ is quaternionic.
\end{lemma}

\begin{proof}
Since $h$ commutes with $\Theta$ by \cref{def:kramers}, the matrix $h$ is quaternionic, and by the closure properties of \cref{sec:kramers}, so are the matrices $e^{th}$ for $t\in\R$, which are limits of real polynomials in $h$, and $\rho$. The matrix $[\,v_i\ \ w_i\,]$ is also quaternionic, because $\Theta v=-\bbJ_N\bar v$ and $\bbJ_N^2=-\I$ give $[\,v_i\ \ w_i\,]J=[\,{-w_i}\ \ v_i\,]=\bbJ_N[\,\bar v_i\ \ \bar w_i\,]$. Since $U_i\ge0$, each $c_k^{1/4}$ is a nonnegative real number, and hence the matrices $R_k^\pm$ are quaternionic as well. Thus the blocks of $G$ in \cref{eq:G} with $k\ne l$ are products of quaternionic matrices. For $k=l=(m,q)$ and $i=i_q$, since $w_i=\Theta v_i$, we have $v_i^\dagger w_i=0$ by \cref{sec:kramers}, so the top-right entry of the middle case equals $v_i^\dagger\rho\,w_i$, and the middle case equals $\sqrt{c_k}\,[\,v_i\ \ w_i\,]^\dagger\rho\,[\,v_i\ \ w_i\,]$, which is a nonnegative multiple of a product of quaternionic matrices. Hence every block of $G$ is quaternionic, and so is $G$.
\end{proof}

\noindent This is the first place in this section where we use the Kramers symmetry of the Hamiltonian $H$ and the attractive sign of its interactions from \cref{def:kramers}.
It is apparent that, from the way we constructed the matrix $G$ and using Wick's theorem, for every $X \subseteq \cK$, by restricting to the principal submatrix of $G$ formed by the block rows and block columns in $X$, we get that
\begin{align}\label{eq:fG}
    f(X) = \cZ_0 \det G[X], \quad \text{and hence} \quad \Tr\left[L^M\right] \,=\, \cZ_0\sum_{X\subseteq\cK}\det G[X].
\end{align}
We also note that, for every $X \subseteq \cK$, the matrix $G[X]$ is quaternionic, and therefore $f(X) \geq 0$ by \cref{lem:kdet}. Let us remark that $f(X)$ is \emph{not} a determinantal point process measure of the matrix $G$, since each element of $X$ selects two rows and two columns to keep in the submatrix $G[X]$. In fact, $G$ is not even guaranteed to be Hermitian.

\paragraph{From counting to sampling.}
Since the weights $\det G[X]$ are nonnegative, we estimate $\Tr[L^M]$ by the standard reduction from counting to sampling \cite{jerrum1986random,jerrum1993ising}. For $t\in[0,1]$, replace every $U_i$ by $tU_i$ which multiplies every $c_k$ by $t$. This replaces $G$ by $\sqrt t\,G$ and $\det G[X]$ by $t^{|X|}\det G[X]$. Let
\begin{align}\label{eq:Wt}
W(t)\,\coloneqq\sum_{X\subseteq\cK}t^{|X|}\det G[X] \quad \text{and} \quad \mu_t(X)\,\coloneqq\,\frac{t^{|X|}\det G[X]}{W(t)}.
\end{align}
With $W(0)=1$ and $W(1)=\Tr[L^M]/\cZ_0$, if we can sample from $\mu_t(X)$ at interaction strengths $0=t_0<t_1<\dots<t_k=1$, we can estimate
\begin{align}\label{eq:telescoping}
 \frac{W(t_{i-1})}{W(t_i)} \,=\, \mathbb E_{X\sim\mu_{t_i}}\left[\left(\frac{t_{i-1}}{t_i}\right)^{|X|}\right] \quad \text{and} \quad \frac{\Tr\left[L^M\right]}{\cZ_0} \,=\, \prod_{i=1}^k\frac{W(t_i)}{W(t_{i-1})}.
\end{align}
Each ratio is thus the average of a function bounded by one, which we can estimate from samples of $\mu_{t_i}$. In \cref{app:counting}, we explain how to choose the strengths $t_i$ and the number of samples required to estimate $\Tr\bigl[L^M\bigr]$ within the required precision.

\subsection{Worm Markov chain}\label{sec:chaindef}
By the end of \cref{sec:green}, it suffices to sample sets $X\subseteq\cK$ with probability proportional to $\det G[X]$ for arbitrary $U_i \geq 0$ from \cref{def:kramers}. To do this, we enlarge the configuration space to include pairs $(A,B)$ of subsets of $\cK$ with $|A|=|B|$ and $\det G[A,B]$, where $A$ selects block rows and $B$ selects block columns of $G$. Note that $G[A,B]$ is quaternionic by \cref{lem:Gquat}, so $\det G[A,B]\ge0$ by \cref{lem:kdet}. For a parameter $0<\lambda<1$, let
\begin{align}\label{eq:soft}
\pi_\lambda(A,B)\,\coloneqq\,\frac{\lambda^{|A\setminus B|}\det G[A,B]}{Z_\lambda} \quad \text{with} \quad Z_\lambda\,\coloneqq\sum_{\substack{A,B\subseteq\cK\\|A|=|B|}}\lambda^{|A\setminus B|}\det G[A,B].
\end{align}
We call the configurations with $A=B$ \emph{valid}. They do not carry any $\lambda$ factor, so conditioned on $A=B$, the distribution $\pi_\lambda$ is the distribution in which $X$ has probability proportional to $\det G[X]$. 

We define the Markov chain starting from $A=B=\varnothing$. At each time step, we stay at the current configuration $(A,B)$ with probability $1/2$. Otherwise, we make a move in two steps.
\begin{enumerate}
\item Delete a block row $k\in A$ or adds a block column $k\notin B$. Since $|A|=|B|$, there are exactly $|\cK|$ such changes, and the chain picks one of them uniformly at random. We accept the change with probability $1$ if $k\in A\setminus B$, and with probability $\lambda$ otherwise. If we rejects the change, the current configuration stays at $(A,B)$.
\item If we accept a change in the previous step, we restore $|A|=|B|$ by adding a block row or deleting a block column, chosen with probability proportional to the determinant of the resulting submatrix of $G$.
\end{enumerate}
The chain satisfies detailed balance with respect to $\pi_\lambda$, so $\pi_\lambda$ is its stationary distribution. Since the chain is lazy, it is also aperiodic.

The enlarged configuration space can be thought of as coming from adding a small pairing field to the Hamiltonian in \cref{def:kramers}. For each interaction $i\in[r]$, let
\begin{align}
\Delta_i\,\coloneqq\,a(w_i)\,a(v_i), \quad \text{so that} \quad n(v_i)\,n(w_i)\,=\,\Delta_i^\dagger\Delta_i,
\end{align}
be the operator that annihilates a pair of fermions in the Kramers partners $v_i$ and $w_i$. Let $\Delta_k\coloneqq\Delta_{i_q}$ for $k=(m,q)\in\cK$, so that $P_k=\Delta_k^\dagger\Delta_k$. Define the Trotter step with a pairing field
\begin{align}\label{eq:trotter-pairing}
L_\lambda\coloneqq C_\lambda C_\lambda^\dagger, \quad \text{where} \quad C_\lambda\coloneqq e^{-\delta H_0/2}\prod_{i=1}^r\left(\I+c_iP_i+\sqrt{\lambda c_i}\,\left(\Delta_i+\Delta_i^\dagger\right)\right).
\end{align}
This is the Trotter step corresponding to the Hamiltonian $H-\sum_ig_i(\Delta_i+\Delta_i^\dagger)$, with pairing field strengths $g_i\coloneqq\sqrt{2\lambda U_i/\delta}$. Expanding $L_\lambda^M$, each insertion $k$ contributes one of $\I$, $c_kP_k$, $\sqrt{\lambda c_k}\,\Delta_k$ and $\sqrt{\lambda c_k}\,\Delta_k^\dagger$. Let $A$ be the set of insertions that annihilate a pair, through $P_k$ or $\Delta_k$, and let $B$ be the set of insertions that create a pair, through $P_k$ or $\Delta_k^\dagger$. Wick's theorem expresses the resulting terms through the same matrix $G$, with the block rows in $A$ and the block columns in $B$. We also get that
\begin{align}\label{eq:pairing-expansion}
\Tr\left[L_\lambda^M\right]\,=\,\cZ_0\,Z_\lambda.
\end{align}
The valid configurations, in which every insertion in $A$ contributes $P_k$, are exactly the terms of the expansion of $\Tr[L^M]$ in \cref{eq:fG}. Hence their probability is
\begin{align}\label{eq:valid-mass}
\pi_\lambda(A=B)\,=\,\frac{\Tr\left[L^M\right]}{\Tr\left[L_\lambda^M\right]},
\end{align}
the ratio of the Trotterized partition functions without and with the pairing field. In \cref{lem:valid}, we show that this ratio is at least $1/2$ when $\lambda$ is inverse polynomially small i.e., when the pairing field is weak.  We defer the proof of fast mixing of the Markov chain to \cref{sec:mixing} where, in \cref{prop:mixing}, we show that the spectral gap of the chain is at least $\lambda/(2|\cK|)$.

\subsection{Estimating the partition function}\label{sec:proofs}

In this subsection, we combine the results so far to prove \cref{thm:Z} and \cref{cor:energies}. By the reduction from counting to sampling at the end of \cref{sec:green}, estimating $\Tr[L^M]$ reduces to sampling sets $X$ with probability proportional to $\det G[X]$. The Markov chain of \cref{sec:chaindef} samples from the enlarged configuration space with a weak pairing field, and conditioned on the valid configurations, its stationary distribution $\pi_\lambda$ is exactly the distribution we want to sample.

It remains to choose a value $\lambda$ so that probablity of \emph{valid} configuration is at least $1/2$ where a smaller value increases the probablity mass of valid configurations.  On the other hand, the spectral gap of the Markov chain is at least $\lambda/(2|\cK|)$ by \cref{prop:mixing}, so a larger $\lambda$ gives a better bound on the mixing time. We therefore choose
\begin{align}
\lambda = \min\left\{\frac12,\ \frac{e^{-\delta\sum_iU_i}}{8M\beta\left(\sum_i\sqrt{U_i}\right)^2}\right\}.
\end{align}

With this choice, the valid configurations have probability at least $1/2$ by \cref{lem:valid}. So we can sample from the distribution of \emph{valid} configurations by running the Markov chain from $(\varnothing,\varnothing)$, keeping the final state if it is valid, and running the chain again otherwise. By \cref{prop:mixing}, each run takes a number of time steps polynomial in $|\cK|/\lambda$ and $\beta\sum_iU_i$, and a constant number of runs suffices on average. The dependence on $\beta\sum_iU_i$ comes from the starting state, whose probability is at least $e^{-\beta\sum_iU_i}/2$ by \cref{lem:valid}. Hence each sample takes polynomial time.

These ingredients give \cref{thm:Z} by standard arguments, as in \cite{bravyi2017ferromagnets,rayudu2025xy}. Let $\varepsilon$ be the target relative error. By Trotter error bounds \cite{childs2021theory}, a polynomially large number $M$ of Trotter steps makes $\Tr[L^M]$ an approximation of $\cZ$ to relative error $\varepsilon/3$. By \cref{lem:counting}, polynomially many samples, at polynomially many interaction strengths $t\le1$, estimate $\Tr[L^M]/\cZ_0$ to relative error $\varepsilon/3$ with probability at least $3/4$. At strength $t$, the weights are those of the same Hamiltonian with every $U_i$ replaced by $tU_i$, so the sampler above provides each of these samples in polynomial time. Finally, \cref{cor:energies} follows from \cref{thm:Z}. The free energy $F=-\beta^{-1}\log\cZ$ inherits an additive error of order $\varepsilon/\beta$. Since $E_0-2N\log2/\beta\le F\le E_0$, $\beta$ of order $N/\varepsilon$ suffices to estimate the ground energy $E_0$.

\section{Fast mixing of the Markov chain}\label{sec:mixing}

In this section, we prove that the spectral gap of the Markov chain defined in \cref{sec:chaindef} is at least $\lambda/(2|\cK|)$. Our proof relies on results from Chen and Liu \cite{chen2026fpras} about sampling from measures whose generating polynomials are Hadamard products of log-concave polynomials. The proof follows in two steps. We first show that, when we set $\lambda = 1$, the weights $\det G[A,B]$ have a strongly log-concave generating polynomial which implies that our Markov chain is fast mixing. To add the $\lambda^{|A\setminus B|}$ tilt to the distribution, we consider the generating polynomial at $\lambda = 1$ and take a Hadamard product with \emph{local} polynomials, \Cref{eq:tilt_polynomials}. We use results from \cite{chen2026fpras} to show that the Markov chain is fast mixing even for the tilted distribution.

\subsection{Log-concave polynomials from quaternionic matrices}
For a function $\mu\colon2^{\cU}\to\R_{\ge0}$ on the subsets of a finite set $\cU$, the generating polynomial of $\mu$ is
\begin{align}
g_\mu(\bx)\coloneqq\sum_{S\subseteq\cU}\mu(S)\,\bx^S,\quad\bx^S\coloneqq\prod_{u\in S}x_u .
\end{align}
We call $g_\mu$ \emph{log-concave} if
\begin{align}
    \nabla^2\log g_\mu\preceq0 \quad \text{on}\quad \R^{\cU}_{>0},
\end{align}
and \emph{strongly log-concave} if every partial derivative $\partial^{\alpha}g_\mu$, of any order, is either identically zero or log-concave on $\R^{\cU}_{>0}$ \cite{anari2024logconcave}. Strong log-concavity of homogeneous $g_\mu$ implies a rapid mixing Markov chain for sampling. Let $\mu$ be a probability distribution on the subsets of size $r$ of $\cU$, and consider the Markov chain that removes a uniformly random element of the current set and then adds an element back, with probability proportional to the weight of the resulting set. In the matroid literature this chain is called the \emph{basis-exchange chain}, and in the high-dimensional expanders literature this is called the \emph{down-up walk}. Anari, Liu, Oveis Gharan and Vinzant \cite{anari2024logconcave} showed that if $g_\mu$ is strongly log-concave, then this chain has spectral gap at least $1/r$. Their main application of this result is an efficient sampler from the uniform distribution on the bases of a matroid, and the first efficient algorithm to approximately count the bases of an arbitrary matroid.

For multiaffine polynomials with nonnegative coefficients log-concavity already implies strong log-concavity. This fact is implicit in \cite{alimohammadi2021fractionally,chen2026fpras}; since neither reference states it in the form we need, we include a short proof.

\begin{lemma}\label{lem:multiaffine}
Let $p$ be a multiaffine polynomial in $n$ variables with nonnegative coefficients. If $p$ is log-concave on $\R^n_{>0}$, then $p$ is strongly log-concave.
\end{lemma}

\begin{proof}
Every partial derivative of a multiaffine polynomial $p$ with nonnegative coefficients is again multiaffine with nonnegative coefficients. So by induction on the order of the derivative it suffices to show that, for each variable $x_u$, the derivative $\partial_up$ is either identically zero or log-concave on $\R^n_{>0}$. Since $p$ is affine in $x_u$, we can write $p=p_0+x_u\,\partial_up$, where neither $p_0$ nor $\partial_up$ depends on $x_u$. Suppose that $\partial_up$ is not identically zero, so that both $p$ and $\partial_up$ are strictly positive on the positive orthant. Writing $x_{-u}$ for the variables other than $x_u$, for every $t>0$ the function $x_{-u}\mapsto\log p(x_{-u},t)-\log t$ is concave on $\R^{n-1}_{>0}$, as it is the restriction of the concave function $\log p$ to the slice $x_u=t$ up to an additive constant, and as $t\to\infty$ it converges pointwise to $\log\partial_up(x_{-u})$. A pointwise limit of concave functions is concave, so $\partial_up$ is log-concave on $\R^{n-1}_{>0}$, and hence on $\R^n_{>0}$, since it does not depend on $x_u$.
\end{proof}

We now show that the polynomial generated by the determinants of square submatrices of a quaternionic matrix $G$ is strongly log-concave. For $\bx\in\R^p$, let $D_{\bx}$ be the $2p\times2p$ block-diagonal matrix with diagonal blocks $x_1\I_2,\dots,x_p\I_2$.

\begin{lemma}\label{lem:Gpoly}
Let $G\in\C^{2p\times2p}$ be a quaternionic matrix, and let
\begin{align}\label{eq:gG}
g_G(\by,\bz) \,\coloneqq \sum_{\substack{A,B\subseteq[p]\\|A|=|B|}}\det G[A,B]\;\bz^{A}\,\by^{[p]\setminus B} \quad \text{for all} \quad \by,\bz\in\R^p.
\end{align}
Then
\begin{align}\label{eq:gG-det}
g_G(\by,\bz) \,=\, \sqrt{\det\left(D_{\by}+G^\dagger D_{\bz}G\right)} \quad \text{for all} \quad \by,\bz\in\R^p_{>0}.
\end{align}
Consequently, $\log g_G$ is concave on $\R^p_{>0}\times\R^p_{>0}$, and $g_G$ is strongly log-concave.
\end{lemma}

\begin{proof}
For $\by,\bz\in\R^p$, let
\begin{align}\label{eq:F}
F(\by,\bz) \,\coloneqq\, \Pf\left(\bbJ_p(D_{\by}+G^\dagger D_{\bz}G)\right).
\end{align}
For $\by,\bz\in\R^p_{\ge0}$, the matrix $D_{\by}+G^\dagger D_{\bz}G$ is Hermitian, positive semidefinite, and quaternionic, so by \cref{eq:P1} we get that
\begin{align}\label{eq:F-sqrt}
F(\by,\bz) \,=\, \sqrt{\det\left(D_{\by}+G^\dagger D_{\bz}G\right)} \quad \text{for all} \quad \by,\bz\in\R^p_{\ge0}.
\end{align}
It remains to show that $F=g_G$. Both $F$ and $g_G$ are homogeneous polynomials of degree $p$ in the $2p$ variables $(\by,\bz)$. It is obvious from the definition of $g_G$ that it is a multiaffine homogeneous polynomial. To see that $F$ is homogeneous, note that $\Pf(tK)=t^p\Pf(K)$ for any skew-symmetric $K\in\C^{2p\times2p}$ and the entries of the matrix in \cref{eq:F} are linear in $(\by,\bz)$. Combining these properties implies that $F(\by,\bz)$ is a homogeneous polynomial with degree $p$.

To prove $F=g_G$, it is enough to show that the coefficients in the polynomials are equal to each other for every monomial. Since both polynomials are homogeneous with degree $p$, every monomial in them contains at most $p$ of the $2p$ variables. Fix $A,B\subseteq[p]$ such that $|A|=|B|$. Since $g_G$ is multiaffine and homogeneous, we can isolate the coefficient $\det G[A,B]$ of the monomial $\bz^A\by^{[p]\setminus B}$ by setting every other variable, apart from the ones in $\bz^A\by^{[p]\setminus B}$, to zero. By doing so, we get that
\begin{align}
g_G(\by_{[p]\setminus B},\bz_A) \,=\, \det G[A,B]\;\bz^A\,\by^{[p]\setminus B},
\end{align}
where we are using the notation $(\by_{[p]\setminus B},\bz_A)$ to indicate that we are setting all the variables $\{y_k : k\in B\}$ and $\{z_k : k\notin A\}$ to zero.

By doing the same in \cref{eq:F-sqrt}, we get that $F(\by_{[p]\setminus B},\bz_A)$ is the square root of $\det(D_{\by_{[p]\setminus B}}+G^\dagger D_{\bz_A}G)$ for all $\by,\bz\in\R^p_{>0}$. To compute this determinant, we use the expansion
\begin{align}\label{eq:det-expansion}
\det\left(D_{\by_{[p]\setminus B}}+G^\dagger D_{\bz_A}G\right) \,=\, \sum_{T\subseteq[2p]}\left(\prod_{i\notin T}\left(D_{\by_{[p]\setminus B}}\right)_{ii}\right)\det\left(G^\dagger D_{\bz_A}G\right)_T,
\end{align}
where $(\,\cdot\,)_T$ denotes the principal submatrix formed by the rows and columns in $T$. This expansion follows from the linearity of the determinant in each row.

Since the diagonal entries of $D_{\by_{[p]\setminus B}}$ vanish in the blocks in $B$, every nonzero term in \cref{eq:det-expansion} has $T$ containing all the rows in block rows of $B$. The matrix $D_{\bz_A}$ has only $2|A|$ nonzero diagonal entries, so $G^\dagger D_{\bz_A}G$ has rank at most $2|A|=2|B|$. Every principal submatrix of $G^\dagger D_{\bz_A}G$ then also has rank at most $2|B|$. So $\det\bigl(G^\dagger D_{\bz_A}G\bigr)_T=0$ whenever $|T|>2|B|$, and every nonzero term in \cref{eq:det-expansion} has $|T|\le2|B|$. So only the term with $T$ equal to the rows in the blocks in $B$ survives, and we get that
\begin{align}
\det\left(D_{\by_{[p]\setminus B}}+G^\dagger D_{\bz_A}G\right)
&\,=\, \left(\by^{[p]\setminus B}\right)^2\det\left(G[A,B]^\dagger D_{\bz}[A]\,G[A,B]\right) \nonumber \\
&\,=\, \left(\bz^A\by^{[p]\setminus B}\right)^2\,\left|\det G[A,B]\right|^2.
\end{align}
Since $\det G[A,B]\ge0$ by \cref{lem:kdet}, we get that $F(\by_{[p]\setminus B},\bz_A)=\det G[A,B]\,\bz^A\by^{[p]\setminus B}=g_G(\by_{[p]\setminus B},\bz_A)$ for all $\by,\bz\in\R^p_{>0}$, and hence as polynomials. This completes the proof that $F=g_G$, and \cref{eq:F-sqrt} gives \cref{eq:gG-det}.

Finally, for $\by,\bz\in\R^p_{>0}$, the matrix $D_{\by}+G^\dagger D_{\bz}G$ is positive definite and linear in $(\by,\bz)$. Since $\log\det$ is concave on positive definite matrices, \cref{eq:gG-det} implies that $\log g_G$ is concave on $\R^p_{>0}\times\R^p_{>0}$. The polynomial $g_G$ is multiaffine, and its coefficients $\det G[A,B]$ are nonnegative by \cref{lem:kdet}, so \cref{lem:multiaffine} shows that $g_G$ is strongly log-concave.
\end{proof}

When we set $\lambda = 1$ in the Markov chain defined in \cref{sec:chaindef}, it exactly corresponds, apart from its laziness, to the basis-exchange chain that samples from submatrices of $G$ with probability proportional to $\det G[A,B]$, i.e., the distribution in \cref{eq:soft} with $\lambda = 1$. By \cref{lem:Gpoly} and the result of Anari, Liu, Oveis Gharan and Vinzant \cite{anari2024logconcave}, this basis-exchange chain mixes rapidly. In the next subsection, we combine this with a result of Chen and Liu \cite{chen2026fpras}, which shows that the Markov chain still mixes rapidly after a $\lambda$-tilt of the distribution.

\subsection{Spectral gap}
We now return to the Markov chain of \cref{sec:chaindef}. Recall from \cref{eq:soft} that the  stationary distribution of the Markov chain is 
\begin{align}\label{eq:soft_1}
\pi_\lambda(A,B)\, \propto\, \lambda^{|A\setminus B|}\det G[A,B]
\end{align}
whose state space is pairs $(A,B)$ of subsets of $\cK$ with $|A|=|B|$, where $A$ selects block rows and $B$ selects block columns of $G$. The valid configurations, with $A=B=X$, do not carry any $\lambda$ factor, so conditioned on $A=B$, the distribution $\pi_\lambda$ is the distribution in which $X$ has probability proportional to $\det G[X]$. At each time step, the chain stays at the current configuration with probability $1/2$. Otherwise, it picks one of the $|\cK|$ changes that delete a block row $k\in A$ or add a block column $k\notin B$ uniformly at random, and accepts the change with probability $1$ if $k\in A\setminus B$, and with probability $\lambda$ otherwise. If it accepts the change, it restores $|A|=|B|$ by adding a block row or deleting a block column, chosen with probability proportional to the determinant of the resulting submatrix of $G$.

\begin{proposition}\label{prop:mixing}
For every $0<\lambda<1$, the Markov chain of \cref{sec:chaindef} has spectral gap at least $\lambda/(2|\cK|)$.
\end{proposition}

\begin{proof}
For $\lambda=1$, the generating polynomial of $\pi_\lambda(A,B)$, after a homogenization step, is exactly equal to $g_{G}(\by, \bz)$ from \cref{eq:gG}, which is strongly log-concave. The Markov chain of \cref{sec:chaindef} is then, apart from its laziness, the basis-exchange chain for a homogeneous strongly log-concave measure, and has spectral gap at least $1/(2|\cK|)$ by the result of \cite{anari2024logconcave} recalled in the previous subsection. For $\lambda<1$, the pairing field tilts the weights by $\lambda^{|A\setminus B|}$. Chen and Liu \cite{chen2026fpras} prove a Poincaré inequality for the basis-exchange chain after a tilt of the stationary distribution, provided that the tilt satisfies a spectral curvature condition.

To make this precise, recall from \cref{eq:gG}, with $[p]$ replaced by $\cK$, that the state $(A,B)$ has the monomial $\bz^A\by^{\cK\setminus B}$ in $g_G$, with coefficient $\det G[A,B]$. The variable $z_k$ appears in this monomial when $k\in A$, and $y_k$ appears when $k\notin B$. Since $|A|=|B|$, the monomial has degree $|\cK|$. This is the homogenization step mentioned above. The first step of the Markov chain in \cref{sec:chaindef} is to remove one of the $|\cK|$ variables of the monomial. Deleting the block row $k\in A$ removes $z_k$, and adding the block column $k\notin B$ removes $y_k$. The second step adds a variable back, with probability proportional to the coefficient of the resulting monomial.

The result of Chen and Liu \cite{chen2026fpras} that we use concerns distributions whose generating polynomial is a Hadamard product, i.e., a coefficient-wise product, of two polynomials, one log-concave and one satisfying a spectral curvature condition. The distribution $\pi_\lambda$ is of this form. Let
\begin{align}\label{eq:tilt_polynomials}
H_\lambda(\by,\bz)\,\coloneqq\,\prod_{k\in\cK}\left(1+y_k+z_k+\lambda\,y_kz_k\right).
\end{align}
The factor of $k$ contributes $\lambda$ to a monomial exactly when both $z_k$ and $y_k$ appear in it, so the coefficient of $\bz^A\by^{\cK\setminus B}$ in $H_\lambda$ is $\lambda^{|A\setminus B|}$ which is exactly what we want. Hence $\pi_\lambda(A,B)$ is proportional to the product of the coefficients of $\bz^A\by^{\cK\setminus B}$ in $g_G$ and in $H_\lambda$.

Consider the Markov chain that removes a variable from the current monomial with probability $c$ times the factor by which this removal changes the coefficient in $H_\lambda$, for some constant $c>0$, and then adds a variable back with probability proportional to the coefficient of the resulting monomial in $g_G$. Since the first step only sees $H_\lambda$ and the second step only sees $g_G$, this chain satisfies detailed balance with respect to the product of the two coefficients. 
Chen and Liu show that, if $g_G$ is log-concave and $H_\lambda$ satisfies a spectral curvature condition, which we state and verify below, then the spectral gap of this chain is at least $c$.\footnote{This follows from \cite[Theorem~1.14]{chen2026fpras} by identifying a monomial with the set of its variables, restricting to monomials of degree $|\cK|$ as in the proof of \cite[Lemma~5.5]{chen2026fpras}, and using the discrete-time form in \cite[Remark~1.9]{chen2026fpras}.}
The Markov chain of \cref{sec:chaindef} is of this form. Removing $z_k$ or $y_k$ multiplies the coefficient in $H_\lambda$ by $1/\lambda$ if $k\in A\setminus B$, and leaves it unchanged otherwise. The chain removes the variable with probability $1/(2|\cK|)$ in the first case and $\lambda/(2|\cK|)$ in the second, so $c=\lambda/(2|\cK|)$.

It remains to state and check the local spectral curvature condition that $H_\lambda$ needs to satisfy. Since $H_\lambda$ is a product of one factor for each $k$, the spectral curvature condition of Chen and Liu \cite[Definition~1.10]{chen2026fpras} splits into a condition on each factor $1+y_k+z_k+\lambda\,y_kz_k$ of $H_\lambda$. The condition is as follows. Removing the variable from the monomial $y_k$ or $z_k$ must not decrease the coefficient. This holds, since the coefficients of $1$, $y_k$ and $z_k$ all equal $1$. Removing $y_k$, $z_k$, or both of them from the monomial $y_kz_k$ multiplies the coefficient by factors $p$, $q$ and $t$, and the spectral curvature condition is that
\begin{align}
\begin{pmatrix} p^2-p & pq-t\\ pq-t & q^2-q\end{pmatrix}\,\succeq\,0.
\end{align}
Here $p=q=t=1/\lambda$, and this matrix is $(1-\lambda)/\lambda^2$ times the all-ones matrix, which is positive semidefinite since $\lambda<1$. In other words, the coefficient $\lambda$ of $y_kz_k$ is at most the product of the coefficients of $y_k$ and $z_k$, so $y_k$ and $z_k$ are negatively dependent in each factor. This completes the proof.
\end{proof}

\section*{Acknowledgments}
\addcontentsline{toc}{section}{Acknowledgments}

C.R.\ acknowledges the funding support by UK Research and Innovation (UKRI) under the UK government's Horizon Europe funding guarantee EP/X032051/1.
J.T.\ was supported by Japan Society for the Promotion of Science (JSPS) KAKENHI Grant Numbers JP25K17310, JP25H01391, JP25H01388, JP25K24845, and JP25K24848. 
O.P.\ was supported by the U.S. Department of Energy, Office of Science, National Quantum Information Science Research Centers, Quantum Systems Accelerator (Award No.\ DE-SCL0000121).
A.Z.\ was supported by the National Nuclear Security Administration’s Advanced Simulation and Computing program. Part of this work was completed while A.Z.\ was supported by the Laboratory Directed Research and Development program at Sandia National Laboratories under the Gil Herrera Fellowship in Quantum Information Science.

This article has been authored by an employee of National Technology \& Engineering Solutions of Sandia, LLC under Contract No.\ DE-NA0003525 with the U.S.\ Department of Energy (DOE). The employee owns all right, title and interest in and to the article and is solely responsible for its contents. The United States Government retains and the publisher, by accepting the article for publication, acknowledges that the United States Government retains a non-exclusive, paid-up, irrevocable, world-wide license to publish or reproduce the published form of this article or allow others to do so, for United States Government purposes. The DOE will provide public access to these results of federally sponsored research in accordance with the DOE Public Access Plan \url{https://www.energy.gov/downloads/doe-public-access-plan}.

\paragraph{AI methodology.}
We used GPT-6 Astra to assist in coming up with the determinantal worm Markov chain and the proof of its fast mixing. This materially affected \cref{sec:chain} and \cref{sec:mixing}. We used GPT-6 Sol to help identify prior literature where worm update rules appeared in fermionic quantum Monte Carlo, discussed in \cref{sec:related_work}. GPT-6 Sol also provided the counterexample in \cref{app:KTR_counterexample}. The authors verified the correctness and originality of all content, including references.

\bibliographystyle{alphaurl}
\bibliography{refs}
\addcontentsline{toc}{section}{References}
\addtocontents{toc}{\protect\setcounter{tocdepth}{1}}

\appendix
\crefalias{section}{appendix}

\section{Lee-Yang property of the attractive Kramers class}\label{app:lee-yang}

We recall the fermionic Lee-Yang class of \cite{rayudu2026fermions} and show that it contains the attractive Kramers class, after a Bogoliubov transformation. For orthonormal vectors $v,w\in\C^{2N}$, define
\begin{align}
q_{v,w}^{\pm}\coloneqq\frac12\bigl(\I-n(v)-n(w)\bigr)
\pm\frac12\bigl(a(v)^\dagger a(w)^\dagger-a(w)a(v)\bigr),
\end{align}
and let $\mathcal Q$ be the set of these operators over all orthonormal pairs and both signs. In particular, $(q_{v,w}^{+})^\dagger=q_{v,w}^{-}$. Pairs defining different elements of $\mathcal Q$ need not be orthogonal to one another.

\begin{definition}[Fermionic Lee-Yang class {\cite{rayudu2026fermions}}]\label{def:fermionic-lee-yang}
A fermionic Lee-Yang Hamiltonian has the form
\begin{align}
H=\ba^\dagger A\ba
+\frac12\sum_{j,k}\left(B_{jk}a_j^\dagger a_k^\dagger+\overline{B_{jk}}a_ka_j\right)
-\sum_\ell\gamma_\ell\left(r_\ell s_\ell+s_\ell^\dagger r_\ell^\dagger\right),
\end{align}
where $A=A^\dagger\succeq0$, $B=-B^\T$, $\gamma_\ell\ge0$, and $r_\ell,s_\ell\in\mathcal Q$.
\end{definition}

This class contains attractive interactions since for any orthonormal pair $v,w$,
\begin{align}
-\left(q_{v,w}^+q_{v,w}^-+q_{v,w}^-q_{v,w}^+\right)
=-2\left(n(v)-\frac12\right)\left(n(w)-\frac12\right)
-\frac{\I}{2}.
\end{align}
The left-hand side is of the form in
\cref{def:fermionic-lee-yang}, obtained by taking
$(r_\ell,s_\ell)=(q_{v,w}^+,q_{v,w}^-)$ and
$(q_{v,w}^-,q_{v,w}^+)$.
Thus arbitrary sums of such attractive interactions are allowed,
including those on overlapping pairs.

The Lee-Yang property concerns the zeros of the partition function as a function of a complex external field. For $\beta>0$, define
\begin{align}
\cZ(z)\coloneqq
\Tr\exp\left[-\beta\left(H+z\cN\right)\right],
\qquad
\cN\coloneqq\sum_{j=1}^{2N}a_j^\dagger a_j,
\qquad h\in\C.
\end{align}
The Lee-Yang theorem of \cite{rayudu2026fermions} states that
\begin{align}
\cZ(z)\ne0
\quad\text{whenever}\quad
\operatorname{Re}z>0
\quad\text{or}\quad
\operatorname{Re}z<-\lambda_{\max}(A).
\end{align}
Thus all zeros lie in the strip
$-\lambda_{\max}(A)\le\operatorname{Re}z\le0$.
In particular, when $A=0$, all zeros lie on the imaginary axis.
The Lee-Yang theorem applies also after a Bogoliubov transformation, provided the number operator field is transformed along with the Hamiltonian.

\begin{proposition}\label{prop:kramers-lee-yang}
Every Hamiltonian in the attractive Kramers class is a fermionic Lee-Yang Hamiltonian after a Bogoliubov transformation with $A=0$.
\end{proposition}

\begin{proof}
Let $H$ be as in \cref{def:kramers}, and work in the Kramers basis of \cref{sec:kramers}. For each Kramers pair $j\in[N]$, define
\begin{align}
b_{2j-1}\coloneqq\frac{a_{2j-1}+a_{2j}^\dagger}{\sqrt2},
\qquad
b_{2j}\coloneqq\frac{a_{2j}-a_{2j-1}^\dagger}{\sqrt2}.
\end{align}
These operators satisfy the canonical anticommutation relations. Since $h$ in \cref{def:kramers} is Hermitian and quaternionic, $B_0\coloneqq-h\bbJ_N$ is antisymmetric. Substituting the inverse transformation into the quadratic term of $H$ gives
\begin{align}
H_0=\ba^\dagger h\ba
=\frac12\Tr h\,\I
+\frac12\sum_{j,k}\left((B_0)_{jk}b_j^\dagger b_k^\dagger
+\overline{(B_0)_{jk}}b_kb_j\right).
\end{align}
Thus the hopping term becomes a pure pairing term, up to an additive constant.

For each interaction, write $n_b(v)\coloneqq b(v)^\dagger b(v)$ and
$\Delta_{b,i}\coloneqq b(w_i)b(v_i)$.
Using the inverse transformation and the canonical anticommutation relations, we obtain
\begin{align}
V_i=-U_i n(v_i)n(w_i)
=-U_i\left(n_b(v_i)-\frac12\right)\left(n_b(w_i)-\frac12\right)
+\frac{U_i}{2}\left(\Delta_{b,i}+\Delta_{b,i}^\dagger\right)
-\frac{U_i}{4}\I.
\end{align}
Hence each interaction becomes a centered attractive interaction together with a pairing term and an additive constant.
\end{proof}

\section{Counting to sampling reduction}\label{app:counting}

In this appendix, we expalin the counting to sampling reduction that we use to estimate the Tortterized partition function. We follow the telescoping argument of \cite[Appendix~B]{rayudu2025xy}, with the interaction strength in place of the inverse temperature. Recall from \cref{eq:Wt} that $W(t)$ is $\Tr[L^M]/\cZ_0$ with every interaction strength $U_i$ replaced by $tU_i$, and that $\mu_t$ is the distribution in which $X$ has probability proportional to $t^{|X|}\det G[X]$. For strengths $0=t_0<t_1<\dots<t_k=1$, \cref{eq:telescoping} gives
\begin{align}
W(1)\,=\,\frac{\Tr\left[L^M\right]}{\cZ_0}\,=\,\prod_{i=1}^k\frac{W(t_i)}{W(t_{i-1})} \quad\text{with}\quad \frac{W(t_{i-1})}{W(t_i)}\,=\,\mathbb E_{X\sim\mu_{t_i}}\left[\left(\frac{t_{i-1}}{t_i}\right)^{|X|}\right].
\end{align}
where we have $W(t_0 = 0) = 1$.

We take $t_1=1/(1+\beta\sum_iU_i)$, and multiply by $1+1/|\cK|$ subsequently at each step until we reaches $t_k=1$. In total, we need
\begin{align}
k\,\le\,2+2|\cK|\log\left(1+\beta\sum_iU_i\right).
\end{align}
number of steps to reach $1$.
This step schedule keeps consecutive partition functions comparable,
\begin{align}\label{eq:ratio-range}
\frac1e\,\le\,\frac{W(t_{i-1})}{W(t_i)}\,\le\,1.
\end{align}
The upper bound holds because every weight $t^{|X|}\det G[X]$ increases with $t$. For the lower bound, we use 
\begin{align}
W(t_i)\,\le\,\left(\frac{t_i}{t_{i-1}}\right)^{|\cK|}W(t_{i-1})\,\le\,e\,W(t_{i-1}) \quad\text{for}\quad i\ge2.
\end{align}

We can estimate each ratio by the average of $(t_{i-1}/t_i)^{|X|}$ over samples from $\mu_{t_i}$, and $W(1)$ by the inverse of the product of these averages. Since each ratio is at least $1/e$, we only need to take number of samples polynomial in $k$ and $1/\varepsilon$ suffices.

\begin{lemma}[Counting to sampling]\label{lem:counting}
Let $0<\varepsilon\le1$ and $n\coloneqq\lceil64ek/\varepsilon^2\rceil$. Suppose that, at each strength $t_i$, the $n$ samples are independent and drawn from a distribution within total variation distance $1/(8kn)$ of $\mu_{t_i}$. Then the estimate of $W(1)$ above has relative error at most $\varepsilon$ with probability at least $3/4$.
\end{lemma}

\begin{proof}
First suppose that the samples are drawn exactly from $\mu_{t_i}$, and let $\hat\theta_i$ be the average at strength $t_i$. The quantity $(t_{i-1}/t_i)^{|X|}$ lies in $[0,1]$, and its mean is at least $1/e$ by \cref{eq:ratio-range}. So the variance of $\hat\theta_i$ is at most $e/n$ times its squared mean. The averages are independent, so their product $\hat\theta\coloneqq\prod_i\hat\theta_i$ satisfies
\begin{align}
\frac{\mathrm{Var}\left(\hat\theta\right)}{\left(\mathbb E\hat\theta\right)^2} \,=\, \prod_{i=1}^k\left(1+\frac{\mathrm{Var}\left(\hat\theta_i\right)}{\left(\mathbb E\hat\theta_i\right)^2}\right)-1 \,\le\, e^{ek/n}-1 \,\le\, \frac{\varepsilon^2}{32},
\end{align}
where the last step uses $n\ge64ek/\varepsilon^2$. By Chebyshev's inequality, $\hat\theta$ is within relative error $\varepsilon/2$ of $\mathbb E\hat\theta=1/W(1)$ with probability at least $7/8$. Since $\varepsilon\le1$, its inverse is then within relative error $\varepsilon$ of $W(1)$. Finally, approximate samples lower this probability by at most $1/8$, since each of the $kn$ samples can be coupled with an exact sample so that the two differ with probability at most $1/(8kn)$.
\end{proof}

\section{Probability of valid configurations}\label{app:valid}

In this appendix, we prove that the valid configurations are not rare when the pairing field is weak. Recall from \cref{eq:valid-mass} that the probability of the valid configurations is the ratio of the Trotterized partition functions without and with the pairing field,
\begin{align}
\pi_\lambda(A=B)\,=\,\frac{\Tr\left[L^M\right]}{\Tr\left[L_\lambda^M\right]},
\end{align}
where $L$ is the Trotter step of \cref{eq:trotter}, and $L_\lambda$ is the Trotter step of \cref{eq:trotter-pairing}, in which every insertion $\I+c_kP_k$ of $L$ becomes $\I+c_kP_k+\sqrt{\lambda c_k}\,(\Delta_k+\Delta_k^\dagger)$. So we need to show that a weak pairing field increases the Trotterized partition function by at most a constant factor.

\begin{lemma}\label{lem:valid}
The Trotterized partition function satisfies $\Tr[L^M]\le e^{\beta\sum_iU_i}\cZ_0$, where $\cZ_0=\Tr[e^{-\beta H_0}]$. Moreover, if
\begin{align}
\lambda\,\le\,\frac{e^{-\delta\sum_iU_i}}{8M\beta\left(\sum_i\sqrt{U_i}\right)^2},
\end{align}
then $\Tr[L_\lambda^M]\le2\Tr[L^M]$, and hence $\pi_\lambda(A=B)\ge1/2$.
\end{lemma}

\begin{proof}
For $0\le i\le r$, let $\Pi_i\coloneqq\prod_{j=i+1}^r(\I+c_jP_j)$, so that $\Pi_r=\I$ and $C=e^{-\delta H_0/2}\Pi_0$ by \cref{eq:trotter}. Each factor $\I+c_jP_j$, where $c_j=\delta U_j/2$ and $P_j$ is a projector, has eigenvalues $1$ and $1+c_j\le e^{c_j}$. Hence $\|\Pi_i^{-1}\|\le1$ and $\|\Pi_i\|\le e^{\delta\sum_jU_j/2}$.

For the first claim, we have
\begin{align}
L\,=\,e^{-\delta H_0/2}\Pi_0\Pi_0^\dagger e^{-\delta H_0/2}\,\preceq\,e^{\delta\sum_iU_i}\,e^{-\delta H_0},
\end{align}
and taking the trace of the $M$th power, with $M\delta=\beta$, gives the first claim.

For the second claim, let
\begin{align}
Q\,\coloneqq\,\prod_{i=1}^r\left(\I+\sqrt{\lambda c_i}\,\Pi_{i-1}^{-1}\left(\Delta_i+\Delta_i^\dagger\right)\Pi_i\right).
\end{align}
Since $\Pi_{i-1}=(\I+c_iP_i)\Pi_i$, the $i$th factor equals $\Pi_{i-1}^{-1}\bigl(\I+c_iP_i+\sqrt{\lambda c_i}\,(\Delta_i+\Delta_i^\dagger)\bigr)\Pi_i$. So the product telescopes, and $CQ=C_\lambda$ by \cref{eq:trotter-pairing}. Hence
\begin{align}
L_\lambda\,=\,CQQ^\dagger C^\dagger\,\preceq\,\|Q\|^2L, \quad\text{so that}\quad \Tr\left[L_\lambda^M\right]\,\le\,\|Q\|^{2M}\,\Tr\left[L^M\right].
\end{align}
The operator $\Delta_i+\Delta_i^\dagger$ has norm one, since $\Delta_i^2=0$ and its square $P_i+\Delta_i\Delta_i^\dagger$ is the projector onto the states in which the modes $v_i$ and $w_i$ are both occupied or both empty. Together with the bounds on $\Pi_{i-1}^{-1}$ and $\Pi_i$, this gives
\begin{align}
\|Q\|^{2M}\,\le\,\prod_{i=1}^r\left(1+\sqrt{\lambda c_i}\,e^{\delta\sum_jU_j/2}\right)^{2M}\,\le\,\exp\left(e^{\delta\sum_jU_j/2}\sqrt{2\lambda M\beta}\,\sum_i\sqrt{U_i}\right),
\end{align}
where we used $1+x\le e^x$, $c_i=\delta U_i/2$ and $M\delta=\beta$. By the assumption on $\lambda$, the exponent is at most $1/2$, so $\Tr[L_\lambda^M]\le e^{1/2}\,\Tr[L^M]\le2\Tr[L^M]$.
\end{proof}

\section{Kramers symmetry and \texorpdfstring{$T$}{T}-invariant decompositions}\label{app:containment}

The purpose of this appendix is to distinguish Kramers symmetry, as we use it in \cref{def:kramers}, from the conventional application of Kramers symmetry in determinant Monte Carlo~\cite{wu2005sufficient} and related methods. The difference is that the latter is a statement about a particular Hubbard-Stratonovich representation of the Trotterized partition function. Thus even if such a representation exists, there is no guarantee that it be efficiently recognizable. Nonetheless, the following restricted ``promise'' definition suffices for our purposes here.

\begin{definition}\label{def:KTR}
    Let $H = H_0 + \sum_i V_i$, where $H_0 = \ba^\dagger h_0\ba$ and each $V_i$ is promised to satisfy the operator identity of the form (for any $\delta > 0$)
    \begin{equation}
        e^{-\delta V_i} = \int d\mu_i(x) \, e^{\ba^\dagger h_i(x) \ba}, \qquad h_i(x) \in \C^{2N \times 2N}
    \end{equation}
    for some positive measure $\mu_i$ such that the integral exists. We say that $H$ exhibits a \emph{$T$-invariant decomposition~\cite{wu2005sufficient}} if there exists an antiunitary operator $\Theta$ with $\Theta^2 = -\I$ such that
    \begin{equation}
        \Theta h_k \Theta^{-1} = h_k \qquad \text{for all } k \geq 0.
    \end{equation}
\end{definition}

The $T$-invariant decomposition is a subset of a broader class of sign-problem-free conditions called Kramers class within Majorana time-reversal symmetry~\cite{li2016majoranatimereversal}, which itself is equivalent to the antisymmetric-$P$ Majorana Kramers positive decomposition~\cite{wei2016majorana}. These are all further subsumed by a generalized semigroup inequality~\cite[Case (f)]{wei2024semigroup}.

On the other hand, our \cref{def:kramers} is strictly about the Hamiltonian terms themselves, which are provided as the input description to the problem. Because we demand efficient recognition of the symmetry, our definition is a strict subset of sign-problem-free Kramers symmetry. We formalize this rigorously below, first showing that \cref{def:kramers} always implies a $T$-invariant decomposition (\cref{app:containment_in_KTR}). Then in \cref{app:KTR_counterexample} we give a simple diagonal counterexample that has a $T$-invariant decomposition, yet cannot be expressed in the form of \cref{def:kramers}.

\subsection{Containment}\label{app:containment_in_KTR}

\begin{proposition}\label{prop:containment_in_KC-MTR}
    Every Hamiltonian $H = H_0 + \sum_i V_i$ as in \cref{def:kramers} exhibits a $T$-invariant decomposition (\cref{def:KTR}).
\end{proposition}

\begin{proof}
    For each $V_i = -U_i n(v_i) n(w_i)$ with the orthogonal pair $v_i, w_i = \Theta v_i$, define $S_i \coloneqq n(v_i) + n(w_i)$. We can use the discrete Hubbard-Stratonovich transformation for attractive interactions~\cite{hirsch1983discrete} to write
    \begin{equation}
        e^{-\delta V_i} = e^{\delta U_i n(v_i) n(w_i)} = \frac{1}{2} \sum_{s\in\{\pm 1\}} e^{(\delta U_i/2 + s\nu_i)(S_i - \I)},
    \end{equation}
    where $\nu_i = \cosh^{-1}(e^{\delta U_i/2})$. Note that the measure here is just the discrete uniform measure over $\{\pm 1\}$. The quadratic generator $S_i - \I$ is Kramers invariant, since
    \begin{equation}
        \Theta_\cF \, n(v_i) \, \Theta_\cF^{-1} = n(w_i), \qquad \Theta_\cF \, n(w_i) \, \Theta_\cF^{-1} = n(v_i),
    \end{equation}
    hence $\Theta_\cF S_i \Theta_\cF^{-1} = S_i$. And by definition, $\Theta_\cF \, H_0 \, \Theta_\cF^{-1} = H_0$.
\end{proof}

\subsection{Counterexample to reverse containment}\label{app:KTR_counterexample}

\begin{proposition}\label{prop:counterexample}
    There exist Hamiltonians exhibiting a $T$-invariant decomposition (\cref{def:KTR}) that cannot be written in the form of \cref{def:kramers}. This is true even with exclusively attractive interactions, arbitrary identity shifts, and arbitrary Bogoliubov transformations.
\end{proposition}

To prove this statement, we will use the following lemma, whose proof is deferred to the end of this appendix. Let ${\bigwedge}^2 \C^4$ be the antisymmetric subspace of two particles in four modes.

\begin{lemma}\label{lem:nonexistence_identity}
The identity operator $\I$ on
${\bigwedge}^2 \C^4$ is not of the form
\begin{equation}\label{eq:cone}
W = \sum_i U_i \ketbra{v_i\wedge\Theta v_i}{v_i\wedge\Theta v_i},
\qquad U_i \geq 0,\quad \|v_i\| = 1,
\end{equation}
for any antiunitary operator $\Theta$ with $\Theta^2=-\I$ on $\C^4$.
\end{lemma}

We prove \cref{prop:counterexample} by exhibiting a counterexample on four modes.

\begin{proof}[Proof of \cref{prop:counterexample}]
    Consider four fermion modes and let $n_p = n(e_p)$, $\cN = \sum_{p=1}^4 n_p$. For a constant $g > 0$ define
    \begin{equation}
        H \coloneqq -g \sum_{1 \leq p < q \leq 4} n_p n_q = -\frac{g}{2} \cN (\cN - \I).
    \end{equation}
    Let $\Theta$ be any antiunitary operator with $\Theta^2 = -\I$. Since $\cN$ commutes with $\Theta$, so too does $H$. This invariance also holds after a Hubbard-Stratonovich transformation~\cite[Eq.~(2)]{wu2005sufficient}:
    \begin{equation}
        e^{-\delta H} = \frac{1}{\sqrt{2\pi}} \int_{-\infty}^\infty dx \, e^{-x^2/2} \exp\mathopen{}\left[ \mathopen{}\left( \sqrt{\delta g} \, x - \frac{\delta g}{2} \right)\mathclose{} \cN \right]\mathclose{}.
    \end{equation}
    The generator of the exponential in the integral is again $\cN$, and so is $\Theta$-invariant. Hence $H$ belongs to \cref{def:KTR}.

    Next we show that $H$ cannot belong to \cref{def:kramers}.
    Suppose there exists some antiunitary $\Theta^2 = -\I$ and some Bogoliubov-transformed modes
    \begin{equation}
        b_p = \sum_{q=1}^4 ( \alpha_{pq} a_q + \beta_{pq} a_q^\dagger ), \qquad \begin{pmatrix}
            \alpha & \beta\\
            \bar{\beta} & \bar{\alpha}
        \end{pmatrix} \text{ is unitary},
    \end{equation}
    such that 
    \begin{equation}\label{eq:H_counterex_form}
        H = \bm{b}^\dagger h \bm{b} - \sum_i U_i n_{\bm{b}}(v_i) n_{\bm{b}}(\Theta v_i), \qquad n_{\bm{b}}(v) = \bm{b}^\dagger vv^\dagger \bm{b}.
    \end{equation}
    We will prove the claim by a contradiction. 

    Let $\cN_{\bm{b}} = \sum_p n_{\bm{b}}(e_p)$ be the number operator in the $\bm{b}$-basis. We can expand in terms of the original $\ba$-basis to get
    \begin{equation}
        \cN_{\bm{b}} = \Tr(\beta^\dagger\beta) + \ba^\dagger A \ba + \frac{1}{2} \mathopen{}\left[ \ba^\dagger D (\ba^\dagger)^\T + \ba^\T D^\dagger \ba \right]\mathclose{}, \qquad A\coloneqq \I - 2\beta^\T\bar{\beta}, \quad D \coloneqq 2\alpha^\dagger\beta.
    \end{equation}
    By definition of the class, $\cN_{\bm{b}}$ and $H$ commute. However in the original basis, $H = -\frac{g}{2} \cN (\cN - \I)$. Let $\ket{\vac}$ be the vacuum in these modes. Then we have both $H\ket{\vac}=0$ and $H a_p^\dagger a_q^\dagger \ket{\vac} = -g \, a_p^\dagger a_q^\dagger \ket{\vac}$. Meanwhile, $\cN_{\bm{b}}$ acting on the same vacuum is
    \begin{equation}\label{eq:Nb_quasiparticle_expansion}
        \cN_{\bm{b}} \ket{\vac} = \Tr(\beta^\dagger\beta) \ket{\vac} + \frac{1}{2} \sum_{p,q} D_{pq} a_p^\dagger a_q^\dagger \ket{\vac}.
    \end{equation}
    Using $[H, \cN_{\bm{b}}] = 0$ we get
    \begin{equation}
        0 = [H, \cN_{\bm{b}}] \ket{\vac} = H\cN_{\bm{b}}\ket{\vac} = -\frac{g}{2} \sum_{p,q} D_{pq} a_p^\dagger a_q^\dagger \ket{\vac}.
    \end{equation}
    The final sum is equivalent to $-g\sum_{p<q} D_{pq} a_p^\dagger a_q^\dagger \ket{\vac}$ since $D = -D^\T$. All terms $a_p^\dagger a_q^\dagger \ket{\vac}$ are linearly independent two-particle basis vectors, so this can be zero only if $D = 0$. 
    
    But $D$ is precisely the component of \cref{eq:Nb_quasiparticle_expansion} that can carry any non-particle-conserving effect in the $\ba$-basis. Thus we must necessarily have $[\cN, \cN_{\bm{b}}] = 0$ for the hypothesis that $H$ be of the form \cref{eq:H_counterex_form} to hold.

    Next, we introduce another set of rotated modes $\bm{d}$ obtained by diagonalizing $\cN_{\bm{b}}$ (hence $A$):
    \begin{equation}\label{eq:Nb_using_dp}
        \cN_{\bm{b}} = \Tr(\beta^\dagger\beta) + \sum_{p=1}^4 \lambda_p(A) d_p^\dagger d_p,
    \end{equation}
    such that
    \begin{equation}\label{eq:bp_using_dp}
        b_p = \sum_{q=1}^4 (V_{pq} d_q + W_{pq} d_q^\dagger) \quad \text{with} \quad \begin{pmatrix}
            V & W\\
            \bar{W} & \bar{V}
        \end{pmatrix} \text{ unitary}.
    \end{equation}
    Now we use the elementary identity $[\cN_{\bm{b}}, b_p] = -b_p$. In terms of the $\bm{d}$-basis, the commutator can also be expressed as
    \begin{equation}
        [\cN_{\bm{b}}, b_p] = \sum_{q=1}^4 (-\lambda_q(A) V_{pq} d_q + \lambda_q(A) W_{pq} d_q^\dagger).
    \end{equation}
    By linear independence of the ladder operators, we must have
    \begin{equation}\label{eq:eigval_A_sign}
        (\lambda_q(A) - 1) V_{pq} = 0, \qquad (\lambda_q(A) + 1) W_{pq} = 0
    \end{equation}
    for every $p, q$. This forces the eigenvalues of $A$ to be $\pm 1$, which can be seen by fixing each $q$ and using the fact that the Bogoliubov transformation must be invertible, hence the $q$th column has at least one nonzero entry at some $p$th row.

    Define particle-hole transformed versions of $\bm{d}$ according to the sign of $\lambda_p(A)$:
    \begin{equation}
        c_p = \begin{cases}
            d_p, & \lambda_p(A) = +1,\\
            d_p^\dagger, & \lambda_p(A) = -1.
        \end{cases}
    \end{equation}
    By \cref{eq:bp_using_dp} and the sign fixing via \cref{eq:eigval_A_sign}, there exists some unitary matrix $R$ that $b_p = \sum_q R_{pq} c_q$. Thus $\cN_{\bm{b}} = \cN_{\bm{c}}$. Recalling the definition of $\cN_{\bm{c}} = \sum_p c_p^\dagger c_p$, we get
    \begin{equation}
        \cN_{\bm{b}} = \sum_{p=1}^4 c_p^\dagger c_p = \sum_{p : \lambda_p(A) = +1} d_p^\dagger d_p + \sum_{p : \lambda_p(A) = -1} (\I - d_p^\dagger d_p) = \kappa + \sum_{p=1}^4 \lambda_p(A) d_p^\dagger d_p,
    \end{equation}
    where $\kappa \coloneqq |\{p : \lambda_p(A) = -1\}|$. From \cref{eq:Nb_using_dp}, this implies that $\Tr(\beta^\dagger\beta) = \kappa$. This establishes the constant term.
    
    For the second term, recall that the modes $\bm{d}$ were obtained by diagonalizing $A$, i.e., $\bm{d} = Q\ba$ for some unitary $Q$. Thus $\cN = \sum_p a_p^\dagger a_p = \sum_p d_p^\dagger d_p$. In terms of the $\bm{c}$-basis,
    \begin{equation}
        \cN = \kappa + \sum_{p=1}^4 \lambda_p(A) c_p^\dagger c_p, \qquad \lambda_p(A) \in \{\pm 1\}.
    \end{equation}
    In terms of the target Bogoliubov basis $\bm{b}$, $\cN = R_\cF \left( \kappa + \sum_{p=1}^4 \lambda_p(A) b_p^\dagger b_p \right) R_\cF^\dagger$ where $R_\cF$ is the many-body operator enacting the particle-preserving rotation $R_\cF^\dagger \, \bm{b} \, R_\cF = R\bm{c}$.

    Finally, we express the Hamiltonian $H = -\frac{g}{2}\cN(\cN-\I)$ in terms of the arbitrary Bogoliubov-rotated basis $\bm{b}$. Write
    \begin{equation}
        H = -\frac{g}{2} R_\cF \, K (K - \I) \, R_\cF^\dagger, \qquad K \coloneqq \kappa + \sum_{p=1}^4 s_p n_{\bm{b}}(e_p)
    \end{equation}
    where $s_p \in \{\pm1\}$. Expand
    \begin{equation}
    \begin{split}
        K(K-\I) &= \left( \kappa + \sum_{p=1}^4 s_p n_{\bm{b}}(e_p) \right) \left( \kappa + \sum_{q=1}^4 s_q n_{\bm{b}}(e_q) - \I \right)\\
        &= \sum_{p,q} s_p s_q n_{\bm{b}}(e_p) n_{\bm{b}}(e_q) + \cdots
    \end{split}
    \end{equation}
    where the omitted terms quadratic and constant terms are all Kramers-invariant. It therefore suffices to consider the quartic component. By hypothesis, we have assumed that the quartic component of $H$ is of the form $V^{(4)} = - \sum_i U_i n_{\bm{b}}(v_i) n_{\bm{b}}(\Theta v_i)$, which is negative semidefinite. Then, first suppose that $s_p \neq s_q$ for some pair $p < q$. This implies that $\ket{\psi_{pq}} \coloneqq R_\cF \, b_p^\dagger b_q^\dagger \ket{\vac_{\bm{b}}}$ is an eigenvector with eigenvalue $+g > 0$:
    \begin{equation}
        V^{(4)} \ket{\psi_{pq}} = -\frac{g}{2} R_\cF \sum_{u,v} s_u s_v n_{\bm{b}}(e_u) n_{\bm{b}}(e_v) R_\cF^\dagger \ket{\psi_{pq}} = -g \, s_p s_q \ket{\psi_{pq}} = g\ket{\psi_{pq}},
    \end{equation}
    which is a contradiction to $V^{(4)} \preceq 0$. Thus all $s_p$ must have the same sign.
    
    Suppose all $s_p s_q = +1$. On the two-particle space ${\bigwedge}^2 \C^4 \cong \spn\{\ket{pq} \coloneqq b_p^\dagger b_q^\dagger \ket{\vac_{\bm{b}}} : p < q\}$, we have for each basis vector
    \begin{equation}
        n_{\bm{b}}(e_u) n_{\bm{b}}(e_v) \ket{pq} = \delta_{(u,v), (p,q)} \ket{pq},
    \end{equation}
    hence $V^{(4)} \ket{\psi_{pq}} = -g\ket{\psi_{pq}}$ for all $p < q$. This implies that $V^{(4)} = -g\I$ when restricted to ${\bigwedge}^2 \C^4$. By \cref{lem:nonexistence_identity}, this cannot be of the form $-\sum_i U_i \ketbra{v_i\wedge\Theta v_i}{v_i\wedge\Theta v_i}$, which is precisely the ${\bigwedge}^2 \C^4$-restriction of $- \sum_i U_i n_{\bm{b}}(v_i) n_{\bm{b}}(\Theta v_i)$. Thus we have a contradiction and $H$ cannot be of the form in \cref{def:kramers}.
\end{proof}

Finally we give the deferred proof of \cref{lem:nonexistence_identity}.

\begin{proof}[Proof of \cref{lem:nonexistence_identity}]
    Choose any normalized Kramers basis $e_1, f_1, e_2, f_2$ where $f_j = \Theta e_j$, and set $\omega \coloneqq (e_1 \wedge f_1 + e_2 \wedge f_2) / \sqrt{2}$. Any unit norm vector of the form $v = \sum_j (\alpha_j e_j + \beta_j f_j)$ has the time reversal $\Theta v = \sum_j(-\bar{\beta}_j e_j+\bar{\alpha}_j f_j)$. Hence
    \begin{equation}
        \braket{\omega | v \wedge \Theta v} = \frac{1}{\sqrt{2}} \sum_{j=1}^2 (|\alpha_j|^2+|\beta_j|^2)=\frac{1}{\sqrt{2}}.
    \end{equation}
    Note that since $v$ has unit norm and is orthogonal to $\Theta v$, their wedge product $v \wedge \Theta v$ also has unit norm. Therefore every $W$ of the form \cref{eq:cone} satisfies
    \begin{equation}
        \braket{\omega | W | \omega} = \frac{1}{2} \sum_i U_i.
    \end{equation}
    Observe that this is also equal to $\frac{1}{2} \Tr W$. If $W$ were equal to $\I$, then $\frac{1}{2} \Tr W = \frac{1}{2} \dim {\bigwedge}^2\C^4 = 3$. However, the left-hand side would evaluate to $\braket{\omega | W | \omega} = \braket{\omega | \omega} = 1$, which is a contradiction.
\end{proof}

\end{document}